%% file: main.tex
\documentclass[sigconf,balance=false]{acmart}

\usepackage{amsmath}
\usepackage{amsthm}

\theoremstyle{definition}
\newtheorem{definition}{Definition}

\newtheorem{theorem}{Theorem}[section]

\newtheorem{lemma}[theorem]{Lemma}

\usepackage{sparklines}
\usepackage[export]{adjustbox}
\usepackage{subfig}
\usepackage{xspace}
\usepackage{xcolor}
\usepackage{soul}
\usepackage{marginnote}
\usepackage{enumitem}
\usepackage[algoruled, linesnumbered, vlined]{algorithm2e}
\usepackage{setspace}
\usepackage{multirow}
\usepackage{multicol}
\usepackage{circledsteps}
\newcommand{\ie}{i.e.,\ }
\newcommand{\eg}{e.g.,\ }

\newcommand{\tinyskip}{\vspace{3pt}}
\newcommand{\mypar}[1]{\tinyskip\noindent\textbf{#1.}\xspace}
\newcommand{\myparnoperiod}[1]{\tinyskip\noindent\textbf{#1}}

\newenvironment{myitemize}{%
\begin{itemize}[leftmargin=1em, itemsep=.1em, parsep=.1em, topsep=.1em,
    partopsep=.1em]}
{\end{itemize}}

\newenvironment{myenumerate}{%
\begin{enumerate}[leftmargin=1em, itemsep=.1em, parsep=.1em, topsep=.1em,
    partopsep=.1em]}
{\end{enumerate}}

\newenvironment{structure*}{\color{blue}\begin{myenumerate}}{\end{myenumerate}}

\sethlcolor{yellow}

\usepackage{popets}

\setcopyright{popets}
\copyrightyear{YYYY}

\acmYear{YYYY}
\acmVolume{YYYY}
\acmNumber{X}
\acmDOI{XXXXXXX.XXXXXXX}
\acmISBN{}
\acmConference{}
\begin{document}

\title{Within-Dataset Disclosure Risk for Differential Privacy}



\author{Zhiru Zhu}
\email{zhiru@uchicago.edu}
\affiliation{%
  \institution{University of Chicago}
  \city{Chicago}
  \state{IL}
  \country{USA}
}

\author{Raul Castro Fernandez}
\email{raulcf@uchicago.edu}
\affiliation{%
  \institution{University of Chicago}
  \city{Chicago}
  \state{IL}
  \country{USA}
}



\input{sections/abstract}


\maketitle

\input{sections/intro}
\input{sections/section2}
\input{sections/section3}

\input{sections/section4}
\input{sections/section5}
\input{sections/evaluation}
\input{sections/related_work}

\input{sections/conclusion}

\bibliographystyle{ACM-Reference-Format}
\bibliography{main}

\input{sections/appendix}

\end{document}

%% file: sections/abstract.tex
\begin{abstract}
Differential privacy (DP) enables private data analysis. In a typical DP deployment, \emph{controllers} manage individuals' sensitive data and are responsible for answering \emph{analysts}' queries while protecting \emph{individuals}' privacy. They do so by choosing the \emph{privacy parameter} $\epsilon$, which controls the degree of privacy for all individuals in all possible datasets. However, it is challenging for controllers to choose $\epsilon$ because of the difficulty of interpreting the privacy implications of such a choice on the \emph{within-dataset} individuals.

To address this challenge, we first derive a \emph{relative disclosure risk indicator} (RDR) that indicates the impact of choosing $\epsilon$ on the within-dataset individuals' disclosure risk. We then design an algorithm to find $\epsilon$ based on controllers' privacy preferences expressed as a \emph{function} of the within-dataset individuals' RDRs, and an alternative algorithm that finds and releases $\epsilon$ while satisfying DP. Lastly, we propose a solution that bounds the total privacy leakage when using the algorithm to answer multiple queries without requiring controllers to set the total privacy budget. We evaluate our contributions through an IRB-approved user study that shows the RDR is useful for helping controllers choose $\epsilon$, and experimental evaluations showing our algorithms are efficient and scalable.
\end{abstract}

%% file: sections/intro.tex
\section{Introduction}


The potential value of releasing statistics from a dataset is offset by the risk of disclosing sensitive information about individuals represented in the dataset. Differential privacy (DP)~\cite{dwork2006differential, dwork2006calibrating} permits releasing statistics while providing theoretical guarantees of privacy, where the degree of privacy is determined by a \emph{privacy parameter} $\epsilon$ --- smaller $\epsilon$ corresponds to a stronger privacy guarantee. While DP presents a rigorous mathematical proof of privacy, choosing $\epsilon$ is challenging in practice since it is difficult to interpret the privacy consequences of making such a choice~\cite{dwork2019differential, cummings2021need, garrido2022lessons, desfontainesblog20240922, smart2022understanding, xiong2020towards, nanayakkara2023chances, dibia2024sok}. In this paper, we propose a method to equip DP practitioners with additional information about the impact of choosing $\epsilon$ on the \emph{within-dataset individuals}' disclosure risk, and we show that this additional information helps controllers choosing $\epsilon$.

In this paper, we consider a private data analysis problem with three types of agents: \emph{analyst}, \emph{controller}, and \emph{individuals}. Each \emph{individual} contributes a data record to a trusted \emph{controller}, who gathers those records into a dataset. \emph{Analysts} want to submit statistical queries over the controller's dataset. The \emph{controller} wants to give \emph{analysts} useful answers to their queries as long as the identities of the \emph{individuals} are protected. To achieve this, the \emph{controller} uses differential privacy, which requires them to choose $\epsilon$ to determine the level of privacy protection offered to \emph{individuals}.

The privacy parameter $\epsilon$ indicates the maximum difference between the probabilities of producing the same query output with or without an arbitrary individual's record, \emph{whether the individual is in the controller's dataset or not}. This maximum difference measures the \emph{worst-case disclosure risk}, which is the probability of identifying an individual given the DP release of a query output. In other words, $\epsilon$ bounds the disclosure risk of \emph{all} individuals across \emph{all} possible datasets. On one hand, this means DP offers a robust privacy guarantee that is independent of any dataset instance. On the other hand, this makes it difficult for controllers to reason about the concrete privacy implications on the \emph{within-dataset} individuals when choosing $\epsilon$. The main challenge of choosing $\epsilon$ is that $\epsilon$ offers a global and probabilistic worst-case measure of disclosure risk, which complicates the controller's decision-making process~\cite{lee2011much}.

In this paper, we provide controllers additional information that complements the worst-case measure of disclosure risk by introducing a \emph{relative, within-dataset} measure of disclosure risk. We call this \textbf{Relative Disclosure Risk Indicator (RDR)}. Given the controller's dataset and analyst's query, the RDR indicates the expected disclosure risk of a specific within-dataset individual \emph{relative} to other within-dataset individuals upon releasing a DP output; this is different from another absolute measure of disclosure risk defined by \citeauthor{lee2011much}~\cite{lee2011much}, which is the maximum probability of identifying \emph{any} within-dataset individuals. With RDR, controllers understand the impact of choosing $\epsilon$ by observing the \emph{distribution} of disclosure risks among the within-dataset individuals. We hypothesize that RDR helps controllers make more informed choices of $\epsilon$ as this additional information empowers controllers to make explicit decisions on their preferred risk distribution over the within-dataset individuals. To test our hypothesis, we run a between-subjects user study where we ask participants to choose $\epsilon$ for a given query and dataset, and we present the treatment group with the additional information of the within-dataset individuals' RDRs. We observe that the treatment group makes more consistent choices of $\epsilon$ compared to the control group, and all participants in the treatment group used RDRs in their decision-making process, thus demonstrating that RDR is effective in helping participants choose $\epsilon$.

Equipped with RDRs, we make several technical contributions. \textbf{First}, we develop the key insight that controllers can express their privacy preferences on the within-dataset individuals as a \emph{function} of the within-dataset individuals' RDRs and a \emph{threshold} over the function output. For instance, controllers can specify the maximum permissible difference between the most at-risk and least at-risk within-dataset individuals. \textbf{Second}, we design an algorithm, called \textsf{Find-$\epsilon$-from-RDR}, that takes controllers' privacy preferences in terms of the within-dataset individuals' RDRs and derives an $\epsilon$ that satisfies those preferences. Because RDRs are data-dependent, they are only visible to controllers and never to analysts, and the derived $\epsilon$ cannot be released to the analysts either. Thus \textbf{third}, we introduce an alternative algorithm, called \textsf{Find-and-release-$\epsilon$-from-RDR}, to enable DP release of the derived $\epsilon$. \textbf{Lastly}, in practice, analysts may submit multiple queries to the controller, and the controller needs to set a \emph{total privacy budget} of their dataset. Choosing this budget suffers from the same difficulty as choosing $\epsilon$~\cite{dwork2019differential, cummings2021need, garrido2022lessons}. We propose an extension to both algorithms that allows controllers to \emph{dynamically} determine whether to allow or deny answering analysts' queries, hence circumventing the need to choose a total budget. We construct a \emph{privacy odometer}~\cite{rogers2016privacy} to provide an end-to-end DP guarantee for our solution.

\mypar{Summary of evaluation results} In our evaluation, we test the hypothesis that the new RDR helps controllers choose $\epsilon$ by conducting an IRB-approved~\footnote{The Institutional Review Board (IRB) is responsible for reviewing studies involving human research subjects and ensuring adequate protection of participants.} user study with 56 participants and the study results demonstrate that participants make more consistent choices of $\epsilon$ when presented with the additional information of within-dataset individuals' RDRs. We show that our proposed algorithms are efficient and scalable; the runtime scales linearly with the dataset size and takes under a few minutes even for one million data records. We conduct ablation studies to analyze the effect of varying controllers' privacy preferences in terms of the within-dataset individuals' RDRs, and we observe that stricter privacy preferences lead to smaller $\epsilon$ chosen.

\mypar{Outline} The rest of the paper is organized as follows. In Section~\ref{sec:background}, we introduce the background on DP and the relevant agents in a standard DP deployment. In Section~\ref{sec:privacyrisk}, we introduce the new RDR definition. In Section~\ref{sec:3derive}, we present the \textsf{Find-$\epsilon$-from-RDR} algorithm. In Section~\ref{sec:algo_ext}, we present the \textsf{Find-and-release-$\epsilon$-from-RDR} algorithm and our solution to answer multiple queries without a total privacy budget while providing end-to-end DP guarantee. We finish with evaluation (Section~\ref{sec:evaluation}), related work (Section~\ref{sec:relatedwork}), and conclusions (Section~\ref{sec:conclusions}).

%% file: sections/section2.tex
\section{Background} \label{sec:background}

In this section, we first introduce the definition of $(\epsilon, \delta)$-differential privacy (Section~\ref{sec:dp_def}), then explain the different agents in a standard DP deployment (Section~\ref{sec:agents}).

\subsection{Differential Privacy} \label{sec:dp_def}

Differential privacy (DP)~\cite{dwork2006differential, dwork2006calibrating} is a mathematical definition of privacy that bounds the effect a single individual's record has on the output of a query. Formally, let the dataset $x = (x_1, \cdots, x_n) \in \mathcal{X}^n$ be a set of $n$ records drawn from a data universe $\mathcal{X}$; we assume that each individual (labeled $1$ to $n$) contributes exactly one record to $x$. A \textit{randomized} mechanism $\mathcal{M}$ satisfies $(\epsilon, \delta)$-differential privacy if for any pair of neighboring datasets $x$ and $x'$ where $x$ and $x'$ differ by a single record, and for all sets of possible output $O \subseteq Range(\mathcal{M})$:
\begin{displaymath}
  \Pr(\mathcal{M}(x) \in O) \leq e^{\epsilon}\Pr(\mathcal{M}(x') \in O) + \delta.
\end{displaymath}
When the \emph{failure parameter} $\delta$ is 0, $\mathcal{M}$ satisfies $\epsilon$-differential privacy. One way of achieving DP is by constructing a mechanism that adds random noise to the query output. The amount of noise is controlled by a tunable \emph{privacy parameter} $\epsilon$. Smaller $\epsilon$ leads to more noise and less accurate output: the choice of $\epsilon$ determines the \emph{privacy-accuracy tradeoff} in DP.

There are two important properties of differential privacy, \emph{post processing} and \emph{sequential composition}, which are building blocks for designing sophisticated mechanisms that satisfy DP.

\mypar{Post processing~\cite{dwork2014algorithmic}} DP is immune to post processing, which means any computation on the output of a DP mechanism will not undermine the DP guarantees. Formally, let $\mathcal{M}$ be an $(\epsilon, \delta)$-DP mechanism, then for any function $f$, $f(\mathcal{M}(\cdot))$ is $(\epsilon, \delta)$-DP.

\mypar{Sequential composition~\cite{dwork2014algorithmic}} Sequentially combining multiple DP mechanisms on the same input dataset results in quantifiable degradation of privacy. Formally, given $k$ independent mechanisms $\mathcal{M}_1, \ldots, \mathcal{M}_k$ that satisfy $(\epsilon_1, \delta_1), \ldots, (\epsilon_k, \delta_k)$-DP respectively, any function $g(\mathcal{M}_1(\cdot), \ldots, \mathcal{M}_k(\cdot))$ is $(\sum_{i=1}^{i=k} \epsilon_i, \sum_{i=1}^{i=k} \delta_i)$-DP.

\subsection{Agents in DP Deployment} \label{sec:agents}

\begin{figure}[h]
    \centering  
    \includegraphics[width=\linewidth]{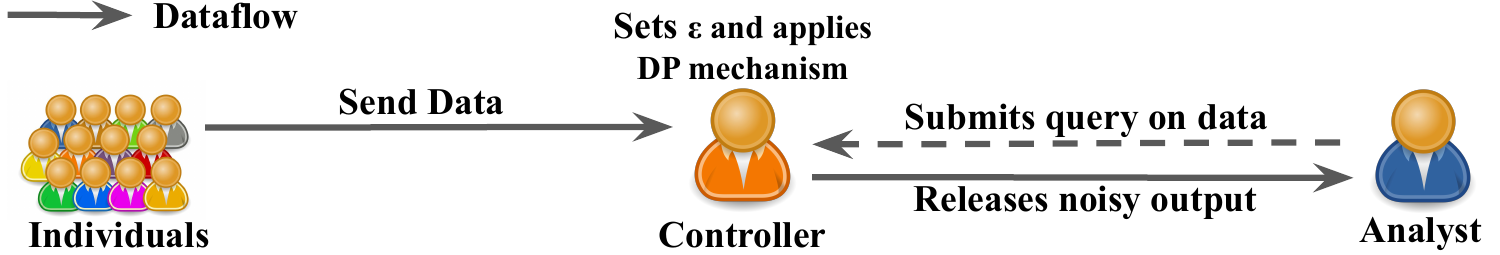}
    \caption{Agents in standard DP deployment}
    \label{fig:agents}
\end{figure}

Figure~\ref{fig:agents} shows a diagram of a standard DP deployment called the \emph{central} DP model~\cite{near2021differential, wagh2021dp}. Specifically, we show the dataflows, \ie how data moves from one agent to another. \emph{Individuals} send their raw data to a \emph{controller}. The \emph{analyst} submits queries to the controller in order to obtain information about the individuals. To protect the privacy of the within-dataset individuals, the controller answers the analyst's queries using a DP mechanism. Crucially, the controller needs to choose the privacy parameter $\epsilon$, which requires them to reason about what is the \emph{worst-case} disclosure risk they are willing to accept not only for the within-dataset individuals, but for all individuals that could be represented in the data universe~\cite{lee2011much}.

One notable example of the central DP model is the application of DP by the US Census in 2020~\cite{abowd2018us}. In this case, the Census Bureau is the controller who has access to individuals' (\ie the US population) data. The analysts' queries are pre-determined by the Census Bureau; the census data is the output of various census queries (\eg population counts in different geographical levels). The Census Bureau made considerable effort to choose the privacy parameter $\epsilon$ for the census queries~\cite{uscensuskeyparam}~\footnote{Technically, the Census Bureau sets the privacy parameter $\rho$ in zero-concentrated differential privacy~\cite{bun2016concentrated}, which can be converted to the $\epsilon$ in $(\epsilon, \delta)$-differential privacy.}.
However, when deploying DP in practice, choosing $\epsilon$ is an extremely challenging task for controllers without the assistance of DP experts~\cite{dwork2019differential, garrido2022lessons, desfontainesblog20240922, Cummings2024Advancing}. 

While there exist other DP deployment models, \ie the local DP~\cite{kasiviswanathan2011can} or shuffler model~\cite{erlingsson2019amplification} where individuals no longer send their raw data directly to the controller, in this paper we focus on addressing the controller's challenges of choosing $\epsilon$ in the central model since this is the most common DP deployment model.

%% file: sections/section3.tex
\section{Relative Disclosure Risk Indicator}
\label{sec:privacyrisk}

We first explain the intuition behind the definition of RDR in Section~\ref{sec:pri_overview}. We then construct \emph{output-dependent} RDR (Section~\ref{sec:oPRI}), which is a relative indicator of the \emph{ex-post per-instance privacy loss}~\cite{redberg2021privately} of each within-dataset individual. Lastly, we propose the RDR definition based on output-dependent RDR (Section~\ref{sec:pri_def}).

\subsection{Why Choosing $\epsilon$ is Hard and RDR Overview} \label{sec:pri_overview}

When controllers decide which $\epsilon$ to choose for an analyst's query, they may naturally ask the question: \emph{how would choosing this $\epsilon$ affect the privacy of the within-dataset individuals I am trying to protect?} However, the value of $\epsilon$ itself does not provide a straightforward answer. Because $\epsilon$ quantifies a data-independent worst-case privacy guarantee, it is challenging to translate this privacy guarantee to a practical measure of disclosure risk of specific within-dataset individuals. The global and probabilistic nature of DP gives a robust privacy guarantee, but it also makes it difficult for controllers to reason about the concrete privacy implications for each within-dataset individual when they apply DP in practice.


We hypothesize that \emph{controllers can only comfortably choose $\epsilon$ if they understand how each within-dataset individual's privacy is affected by that choice}. A fundamental challenge in deploying differential privacy is translating the abstract guarantee of $\epsilon$ into a concrete measure of disclosure risk. \citeauthor{lee2011much} address this challenge by defining disclosure risk as the maximum posterior probability of re-identifying \emph{any} within-dataset individuals~\cite{lee2011much}. This provides a single, \emph{absolute} measure of the worst-case disclosure risk, establishing a ceiling on the harm to the single most vulnerable within-dataset individual. While valuable, this single-point estimate conceals the underlying \emph{distribution} of risk below that ceiling. A controller informed only of the maximum disclosure risk of the within-dataset individuals cannot distinguish between the scenario where the risk is borne equally by all individuals and one where it is dangerously concentrated on a vulnerable few. Therefore, our goal is to design a \emph{relative, per-individual} indicator that complements this absolute, worst-case measure of disclosure risk; we call it the \textbf{Relative Disclosure Risk Indicator (RDR)}. Crucially, the RDR gives controllers a view into the risk distribution of the within-dataset individuals. For instance, by comparing the RDRs of the most and least at-risk individuals, or by examining the overall variance of the RDR distribution, a controller can make a more informed judgment about whether a given $\epsilon$ provides an equitable level of protection across the within-dataset individuals. The RDR complements any information controllers already have access to, and it remains private information to controllers only.


\subsection{Output-dependent RDR} \label{sec:oPRI}

We first construct \emph{output-dependent} relative disclosure risk indicator (oRDR) as the basis for the generic RDR definition. We denote an $(\epsilon, \delta)$-DP mechanism as $\mathcal{M}: \mathcal{X}^n \times \mathcal{Q} \rightarrow \mathcal{Y}$, where $\mathcal{Q}$ is the query space and $\mathcal{Y}$ is the output space of $\mathcal{M}$. Specifically, we consider numeric query $q: \mathcal{X}^n \rightarrow \mathbb{R}^k$ that maps a dataset to $k$ real numbers. Given a query $q \in \mathcal{Q}$ and a query output $o \sim \mathcal{M}(x, q, \epsilon, \delta)$, we define the \emph{output-dependent} RDR of an individual $i \in [n]$ as follows: 
\begin{definition}[\textbf{Output-dependent Relative Disclosure Risk Indicator}]
The output-dependent relative disclosure risk indicator of an individual $i$ parameterized by output $o$ is defined as:
    \[    
    RDR_i^{(o)} = \lVert o - q(x_{-i}) \rVert_p,
    \]
    where $x_{-i}$ is the neighboring dataset of $x$ without $i$'s record $x_i$. The value of $p$ depends on the specific mechanism $\mathcal{M}$ being applied: if $\mathcal{M}$ is the \emph{Laplace Mechanism}~\cite{dwork2014algorithmic}, $p=1$ (\ie the $\ell_1$ norm); if $\mathcal{M}$ is the \emph{Gaussian Mechanism}~\cite{dwork2014algorithmic}, $p=2$ (\ie the $\ell_2$ norm).
\end{definition}

Next, we show the connection between output-dependent RDR and \emph{ex-post per-instance privacy loss}~\cite{redberg2021privately}. Specifically, we show that when Laplace Mechanism~\cite{dwork2014algorithmic} or Gaussian Mechanism~\cite{dwork2014algorithmic} is applied to compute the query output $o$, $RDR_i^{(o)}$ becomes a \emph{relative} indicator of ex-post per-instance privacy loss associated with a within-dataset individual $i$.

\begin{definition}[\textbf{Ex-post per-instance privacy loss}~\cite{redberg2021privately}]
    For a DP mechanism $\mathcal{M}$ and a query $q \in \mathcal{Q}$, the ex-post per-instance privacy loss given a realized query output $o \sim \mathcal{Y}$, a fixed dataset $x$, and an individual's record $x_i \in x$ is
    \[
    \epsilon(o, x, x_{-i}) = \ln \left( \frac{\Pr[\mathcal{M}(x, q, \epsilon, \delta) = o]}{\Pr[\mathcal{M}(x_{-i}, q, \epsilon, \delta) = o]} \right).
    \]
\end{definition}


Note the difference between $\epsilon$, which is an input to a DP mechanism, and $\epsilon(\cdot)$, which is a function that denotes the privacy loss conditioned on a given input. Intuitively, using ex-post per-instance privacy loss answers the question: \emph{what is the realized privacy loss incurred to each within-dataset individual when $o$ is released?} Hence, it is a fine-grained characterization of the privacy loss defined in standard DP~\cite{dwork2014algorithmic}, which is the worst-case bound for all individuals in all possible datasets. 
We derive the output-dependent RDR from the ex-post per-instance privacy loss so that it describes the disclosure risk of a within-dataset individual \emph{relative} to other within-dataset individuals. Next, we explain the connection between oRDR and ex-post per-instance privacy loss when the underlying DP mechanism $\mathcal{M}$ is Laplace Mechanism or Gaussian Mechanism.      

\begin{definition}[\textbf{Laplace Mechanism}~\cite{dwork2014algorithmic}]\label{def:laplace}
    Given query $q: \mathcal{X}^n \rightarrow \mathbb{R}^k$, dataset $x \in \mathcal{X}^n$, $\epsilon > 0$, Laplace Mechanism is defined as:
    \[
    \mathcal{M}_{Lap}(x, q, \epsilon) = q(x) + (Z_1, \ldots, Z_k),
    \]
    where $Z_i$ are i.i.d random variables drawn from the Laplace distribution with probability density function $p_b(z) = \frac{1}{2b} \exp (-|z|/b)$ and the scale $b = \frac{\Delta_1}{\epsilon}$~\footnote{The global sensitivity of q in the $\ell_1$ norm is defined as $\Delta_1 = \max \lVert q(x) - q(x') \rVert_1$, where $x$ and $x'$ are neighboring datasets in $\mathcal{X}^n$.}. The Laplace Mechanism satisfies $\epsilon$-DP.
\end{definition}

\myparnoperiod{Claim:} For all individuals $i, j$ where $i \neq j$, if $RDR_i^{(o)} > RDR_j^{(o)}$ and $o$ is computed using Laplace Mechanism, then
\[
\epsilon_{Lap}(o, x, x_{-i}) > \epsilon_{Lap}(o, x, x_{-j})
\]

\myparnoperiod{Proof:} The ex-post per-instance privacy loss under Laplace Mechanism is computed as:
\begin{align*}
     \epsilon_{Lap}(o, x, x_{-i}) &= \ln \left( \prod_{j = 1}^{k} \frac{\exp (-\epsilon |o_j - q(x)_j| / \Delta_1)}{\exp (-\epsilon |o_j - q(x_{-i})_j| / \Delta_1)} \right) \\ &= \epsilon \frac{\lVert o - q(x_{-i}) \rVert_1 - \lVert o - q(x) \rVert_1}{\Delta_1}.
\end{align*}

$RDR_i^{(o)} > RDR_j^{(o)}$ means $\lVert o - q(x_{-i}) \rVert_1$ > $\lVert o - q(x_{-j}) \rVert_1$ (the oRDR is in $\ell_1$ norm since Laplace Mechanism is used), thus \\ $\epsilon_{Lap}(o, x, x_{-i}) > \epsilon_{Lap}(o, x, x_{-j})$. 


\begin{definition}[\textbf{Gaussian Mechanism}~\cite{dwork2014algorithmic}]\label{def:gaussian}
    Given query $q: \mathcal{X}^n \rightarrow \mathbb{R}^k$, dataset $x \in \mathcal{X}^n$, $\epsilon, \delta > 0$, Gaussian Mechanism is defined as:
    \[
    \mathcal{M}_{Gau}(x, q, \epsilon, \delta) = q(x) + (Z_1, \ldots, Z_k),
    \]
    where $Z_i$ are i.i.d random variables drawn from the Gaussian distribution $\mathcal{N}(\sigma^2)$ with center $0$ and variance $\sigma^2 = \frac{2 \Delta_2^2 \ln(1.25/\delta)}{\epsilon^2}$~\footnote{The global sensitivity of q in the $\ell_2$ norm is defined as $\Delta_2 = \max \lVert q(x) - q(x') \rVert_2$, where $x$ and $x'$ are neighboring datasets in $\mathcal{X}^n$.}. The Gaussian Mechanism satisfies $(\epsilon, \delta)$-DP.
\end{definition}

\myparnoperiod{Claim:} For all individuals $i, j$ where $i \neq j$, if $RDR_i^{(o)} > RDR_j^{(o)}$ and $o$ is computed using Gaussian Mechanism, then
\[
\epsilon_{Gau}(o, x, x_{-i}) > \epsilon_{Gau}(o, x, x_{-j})
\]

\myparnoperiod{Proof:} The ex-post per-instance privacy loss under Gaussian Mechanism is computed as:
\begin{align*}
     \epsilon_{Gau}(o, x, x_{-i}) &= \ln \left( \frac{\exp (- \lVert o - q(x) \rVert_2^2 / 2\sigma^2)}{\exp (- \lVert o - q(x_{-i}) \rVert_2^2 / 2\sigma^2)} \right) \\ &=  \frac{\lVert o - q(x_{-i}) \rVert_2^2 - \lVert o - q(x) \rVert_2^2}{2\sigma^2}.
\end{align*}

$RDR_i^{(o)} > RDR_j^{(o)}$ means $\lVert o - q(x_{-i}) \rVert_2$ > $\lVert o - q(x_{-j}) \rVert_2$; the oRDR is in $\ell_2$ norm under Gaussian Mechanism.

\subsection{RDR: Formal Definition} \label{sec:pri_def}

Output-dependent RDR allows controllers to observe the relative disclosure risk of the within-dataset individuals when releasing a specific DP output, but not when choosing $\epsilon$ since the same DP output could be produced by different $\epsilon$. Building from the definition of oRDR, we propose a generic definition of RDR that indicates the relative disclosure risk of the within-dataset individuals \emph{in expectation} across all possible outputs that can be generated by a DP mechanism with a specific $\epsilon$.  

\begin{definition}[\textbf{Relative Disclosure Risk Indicator}]
    Given DP mechanism $\mathcal{M}$, query $q: \mathcal{X}^n \rightarrow \mathbb{R}^k$, dataset $x \in \mathcal{X}^n$, and parameters $\epsilon, \delta$, we define the relative disclosure risk indicator of an individual $i \in [n]$ as:
    \[
    RDR_i = \mathbb{E}_{o \sim \mathcal{M}(x, q, \epsilon, \delta)}[RDR_i^{(o)}].
    \]
\end{definition}

Next, we derive an upper bound of the RDR under Laplace Mechanism and Gaussian Mechanism, since it is easier to compute than the exact RDR value and suffices for interpretation. We will refer to this upper bound as  RDR in the rest of the paper.

\begin{lemma}\label{lemma:pri_lap} $RDR_i$ upper bound under Laplace Mechanism is
    \[
      \lVert q(x) - q(x_{-i}) \rVert_1 + k \Delta_1 / \epsilon.
    \]
\end{lemma}

\begin{lemma}\label{lemma:pri_gau} $RDR_i$ upper bound under Gaussian Mechanism is
    \[
     \sqrt{\lVert q(x) - q(x_{-i}) \rVert_2^2 + k \sigma^2}.
    \]
\end{lemma}

$k$ is the output dimension of query $q$. As noted in Definition \ref{def:laplace} and \ref{def:gaussian}, $\Delta_1 / \epsilon$ is the noise scale of Laplace distribution, and $\sigma^2$ is the variance of Gaussian distribution. 
We omit details of the calculation due to space constraints and include them in the Appendix.

\mypar{Interpretation of RDR} The value $\lVert q(x) - q(x_{-i}) \rVert_p$ is defined as the \emph{per-instance sensitivity}~\cite{wang2019per} of query $q$ given a fixed $x$ and $x_{-i}$, where $p$ denotes the $\ell_p$ norm. Intuitively, per-instance sensitivity signals how \emph{sensitive} a within-dataset individual's record is; if removing an individual's record causes large differences in the query output, then this individual is more \emph{distinguishable} than others and more at risk of being identified. When no DP noise is added to the query output (\ie as $\epsilon$ approaches infinity), for both RDR values in Lemma~\ref{lemma:pri_lap} and Lemma~\ref{lemma:pri_gau}, the noise component (\ie $k \Delta_1 / \epsilon$ and $k \sigma^2$) approaches 0 and RDR is reduced to per-instance sensitivity, thus indicating a within-dataset individual's relative disclosure risk if the raw query output was released. As $\epsilon$ gets smaller, the noise component becomes larger; this means a within-dataset individual with large per-instance sensitivity will be able to "blend-in" with others if their RDRs become dominated by the noise component. On the other end of the spectrum, as $\epsilon$ approaches 0, the RDR of every within-dataset individual will all be infinitely large, thus \emph{no within-dataset individual is at more risk than others} (note that RDR is a relative indicator). We now make a few remarks:

\mypar{RDR is only meaningful when compared to other RDRs} The value of a within-dataset individual's RDR alone does \emph{not} indicate the absolute disclosure risk of the individual. A within-dataset individual's RDR is only meaningful when compared to others in the same dataset as it indicates whether they are more or less at risk \emph{relative} to others. Hence when $\epsilon$ approaches 0 and all RDRs are infinitely large, every within-dataset individual becomes \emph{indistinguishable} from every other, and they share the strongest privacy.


\mypar{RDR is data-dependent thus leaks information about within-dataset individuals} Controllers already have access to individuals' raw data, and they can compute the RDRs of the within-dataset individuals from the raw data. The RDR is private information available to controllers \emph{only} and should never be shared with others.



%% file: sections/section4.tex
\section{Deriving Epsilon based on RDR}
\label{sec:3derive}

In this section, we describe the key idea of using RDR to express controllers' privacy preferences over the within-dataset individuals (Section~\ref{sec:3.1equal}), then introduce an algorithm to choose $\epsilon$ by using RDR as a more intuitive proxy (Section~\ref{sec:3.2algo}). 

\subsection{Using RDR to Express Privacy Preferences} \label{sec:3.1equal}

The RDR complements the worst-case bound of disclosure risk given by $\epsilon$ and becomes an additional indicator of relative disclosure risk associated with each within-dataset individual. A controller may express their privacy preference over the within-dataset individuals in terms of their RDRs, and choose an $\epsilon$ such that the corresponding RDRs satisfy their privacy preference. Next, we convey this intuition through an example. 


\begin{figure}[h]
    \centering  
    \includegraphics[width=\linewidth]{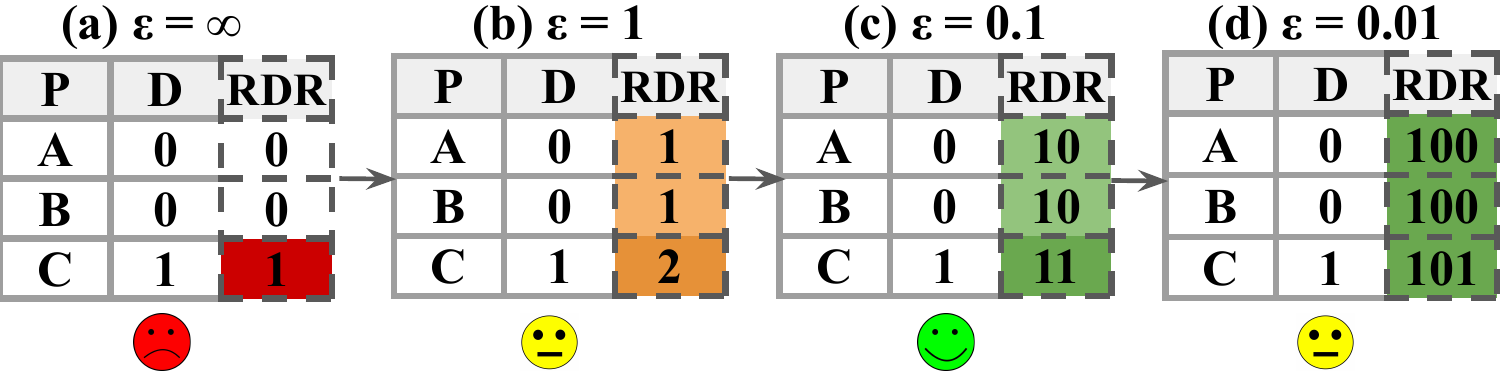}
    \caption{Example of using RDR to find $\epsilon$. The query is to count the number of patients (column \textsc{P}) who have a certain disease (column \textsc{D}) and Laplace Mechanism is used to compute the query. For each $\epsilon$, we show the corresponding RDR of each within-dataset patient.}
    \label{fig:pri}
\end{figure}

\mypar{Example illustrating one use of the RDR} Figure \ref{fig:pri} shows a dataset of three records, where each record consists of a patient's ID (column \textsc{P}) and whether the patient has a certain disease (column \textsc{D}); the value 1 means the patient has the disease and 0 otherwise. Consider an analyst issuing a query about the number of patients who have the disease. To protect the privacy of the within-dataset patients, the controller decides to use the Laplace Mechanism to compute a noisy count. The controller does not know which $\epsilon$ they should use, so they first select a set of candidate $\epsilon$ values (\eg $\infty, 1, 0.1, 0.01$) and compute the associated RDR of each within-dataset patient (shown as an additional column by the dataset).  

When $\epsilon = \infty$, \ie when no DP is applied, the controller observes that patient C who is the only one with the disease will have an RDR of 1, and others will have an RDR of 0, which indicates patient C has significantly greater disclosure risk than others (as discussed in Section~\ref{sec:pri_def}, when $\epsilon = \infty$, RDR is reduced to per-instance sensitivity). By decreasing $\epsilon$, the resulting RDRs become \emph{closer} across the within-dataset patients. The RDR is computed by adding $\frac{1}{\epsilon}$ to each within-dataset patient's per-instance sensitivity since the Laplace Mechanism is used and the query's sensitivity $\Delta_1 = 1$ (Lemma~\ref{lemma:pri_lap}). Consider the range of RDRs for each $\epsilon$: even though the difference between the minimum RDR and maximum RDR of the within-dataset patients is always 1 (this only holds for our specific example~\footnote{In fact, given a fixed dataset, query, DP mechanism, for any $\epsilon$, the maximum difference among the within-dataset individuals' RDRs is the query's local sensitivity~\cite{nissim2007smooth}.}), the \emph{ratio} between the minimum and maximum RDR becomes closer to 1 as $\epsilon$ gets smaller. In other words, the within-dataset patients' RDRs become relatively closer to each other, and specifically patient C becomes less distinguishable. 

The controller chooses $\epsilon$ based on their privacy preference in terms of the ratio between the maximum and maximum RDR of the within-dataset patients. Between $\epsilon = 1$ and $\epsilon = 0.1$, the controller might be more comfortable choosing $\epsilon = 0.1$ since the within-dataset patients' RDRs are closer together (the ratio between the minimum and maximum RDR increases to $\frac{10}{11}$ from $\frac{1}{2}$). Between $\epsilon = 0.1$ and $\epsilon = 0.01$, although the latter produces even closer RDRs (the ratio becomes $\frac{100}{101}$), it comes at the cost of worse output accuracy. Thus one sensible choice is to choose $\epsilon = 0.1$ as it yields higher accuracy for the analyst receiving the output.    

\mypar{Key idea} When a controller computes a query by applying a DP mechanism with a specific $\epsilon$, observing the RDRs associated with the within-dataset individuals reveals the distribution of the individuals' disclosure risks. This enables the controller to formally express their privacy preference over the within-dataset individuals by specifying a \emph{preference function} $f: \mathbb{R}^k \rightarrow \mathbb{R}$ in terms of individuals' RDRs, \eg $f(RDR) = RDR_{min} / RDR_{max}$, and a \emph{threshold} $\tau \in \mathbb{R}$ over the output of $f(RDR)$, \eg a percentage $0 \le \tau_p \le 1$. The controller's privacy preference are satisfied when $f(RDR) \ge \tau$. In the above example, if the controller sets $\tau_p = 0.9$, their privacy preference are satisfied when $\epsilon = 0.1$ since $\frac{10}{11} \ge 0.9$.

\subsection{The \textsf{Find-$\epsilon$-from-RDR} Algorithm} 
\label{sec:3.2algo}

We design the \textsf{Find-$\epsilon$-from-RDR} algorithm (Algorithm~\ref{alg:find_eps}) to operationalize the key idea presented above. The algorithm provides a general framework for finding a suitable $\epsilon$ that aligns with a controller's specific privacy preference on the within-dataset individuals. Instead of prescribing a single, fixed RDR metric, our approach allows the controller to define their own \emph{preference function} over RDRs, $f(RDR)$, and a corresponding threshold, $\tau$. The algorithm then finds the \emph{largest} candidate $\epsilon$ (preferring higher utility) that satisfies the controller's stated preference, $f(RDR) \ge \tau$.

\begin{algorithm}[h]
\scriptsize
\SetKw{Continue}{continue}
\SetKwInOut{Input}{Input}
\SetKwInOut{Output}{Output}
\Input{DP mechanism $\mathcal{M}$, dataset $x$, query $q$, preference function $f$, threshold $\tau$, set of candidate $\epsilon$ values $\mathcal{E}$, failure parameter $\delta$} 
\Output{DP output $o$}
\BlankLine

$x' \leftarrow \textsc{project-query-attributes}(x, q)$ \label{algo1:project-attr} \\

$PIS \leftarrow map()$ \label{algo1:create-map} \\ 

\For{$x_i \in x'$}{
    \If{$x_i \notin PIS$}{
        $PIS[x_i] \leftarrow \textsc{compute-per-instance-sensitivity}(x', x_i, q, M)$
    } 
} \label{algo1:compute-pis}

\tcc*[h]{{\scriptsize Controllers may observe the computed per-instance sensitivity of each individual first, and set $f$ and $\tau$ based on their privacy preferences}}\\

\For{$\epsilon \in \mathcal{E}$}{ \label{algo1:loop}

    $RDR \leftarrow list()$ \label{algo1:create-list} \\
    \For{$x_i \in x'$}{
        $RDR_i \leftarrow \textsc{compute-rdr}(PIS[x_i], M, q, \epsilon, \delta)$ \\
        $RDR.add(RDR_i)$ \label{algo1:compute-rdr}
    }
    

    \tcc*[h]{{\scriptsize Check whether $f(RDR)$ satisfy controller's privacy preference}}\\
    \If{$f(RDR) \ge \tau$}{ \label{algo1:check-rdr}

        \tcc*[h]{{\scriptsize Found suitable $\epsilon$. Compute DP output using chosen $\epsilon$}}\\
        $o \leftarrow \mathcal{M}(x', q, \epsilon, \delta)$ \label{algo1:compute-dp-output} \\
        \Return $o$ \label{algo1:return}
    }
}
\Return nil
\caption{Find-$\epsilon$-from-RDR}
\label{alg:find_eps}
\end{algorithm}

\mypar{Choosing preference function} The central argument of our work is that the controller, who has complete knowledge of individuals' data and is responsible for protecting those within-dataset individuals, is the best-positioned agent to define what constitutes an acceptable risk distribution. By first computing and observing the within-dataset individuals' per-instance sensitivities (Algorithm~\ref{alg:find_eps}, lines \ref{algo1:create-map} - \ref{algo1:compute-pis}, which will be used to compute individuals' RDRs later), the controller gains a concrete understanding of the baseline disclosure risk of each individual. Armed with this information, they can formulate a preference $f, \tau$ that is not abstract, but rather grounded in the specific characteristics of their data. This reframes the problem from the difficult task of interpreting a global, worst case $\epsilon$ to the more intuitive one of expressing a direct policy on the relative risks they are willing to accept for the individuals they protect.

The flexibility of the preference function is a key strength of our framework, allowing controllers to express a wide variety of privacy preferences. As shown in our previous example, a controller may wish to ensure that no single individual is disproportionately more at risk than any other. This can be expressed using the preference function $f(RDR) = RDR_{min} / RDR_{max} \ge \tau_p$, where $\tau_p$ is a value close to 1 (\eg 0.95). Alternatively, a controller might be less concerned with outliers and more with ensuring the overall risk distribution is tight. This goal can be captured by minimizing variance: $f(RDR) = -Var(RDR) \ge -\tau_{var}$ (equivalent to $Var(RDR) \le \tau_{var}$), for a small variance threshold $\tau_{var}$. In addition, a controller may be particularly concerned about protecting vulnerable subgroups (\eg minorities in a demographic dataset). They could define a preference that compares the median risk of a minority group to that of the majority group: $f(RDR) = RDR_{median\_majority} / RDR_{median\_minority} \ge \tau_g$, ensuring the risks between groups remain balanced. The choice of the preference function $f$ and threshold $\tau$ is not an arbitrary task but a deliberative, socio-technical process that depends on  the controller's privacy goals and their understanding of the data, now complemented with the additional information of the per-instance sensitivities and RDRs of within-dataset individuals.

\mypar{Detailed Algorithm} Algorithm \ref{alg:find_eps} shows the pseudo-code for finding $\epsilon$ from RDR and releasing DP output $o$. The input includes the set of candidate $\epsilon$ values to choose from; for example, the values could range from 10 to 0.01 in a fixed interval. The algorithm first projects relevant query attributes from the dataset (line \ref{algo1:project-attr}) to avoid unnecessary recomputation of duplicate data (\eg if the query only involves the \texttt{age} attribute, the algorithm only needs to compute the RDRs of hundreds of unique ages, regardless of the original size of the dataset). Then, for each unique within-dataset individual's data $x_{i}$, the algorithm computes its per-instance sensitivity, which depends on the query and which DP mechanism will be used (line \ref{algo1:create-map} - \ref{algo1:compute-pis}); the per-instance sensitivity will be used to compute the RDR of each within-dataset individual. At this point, the controller may observe each within-dataset individual's per-instance sensitivity and sets the preference function $f$ and corresponding threshold $\tau$ based on their privacy preference over the within-dataset individuals.

Next, the algorithm iterates over the set of candidate $\epsilon$ values in descending order (line \ref{algo1:loop}). For each $\epsilon$, the algorithm computes the RDR of each individual (line \ref{algo1:create-list} - \ref{algo1:compute-rdr}). If $f(RDR) \ge \tau$, then the algorithm has found a suitable $\epsilon$ that satisfies the controller's privacy preference, so it computes a DP output using the chosen $\epsilon$ and returns the output (line \ref{algo1:compute-dp-output} - \ref{algo1:return}). Note that the algorithm does \emph{not} release the chosen $\epsilon$, since $\epsilon$ is chosen based on data-dependent RDRs and only the controller can access the RDRs and chosen $\epsilon$.

\begin{figure}[h]
    \centering  
    \includegraphics[width=\linewidth]{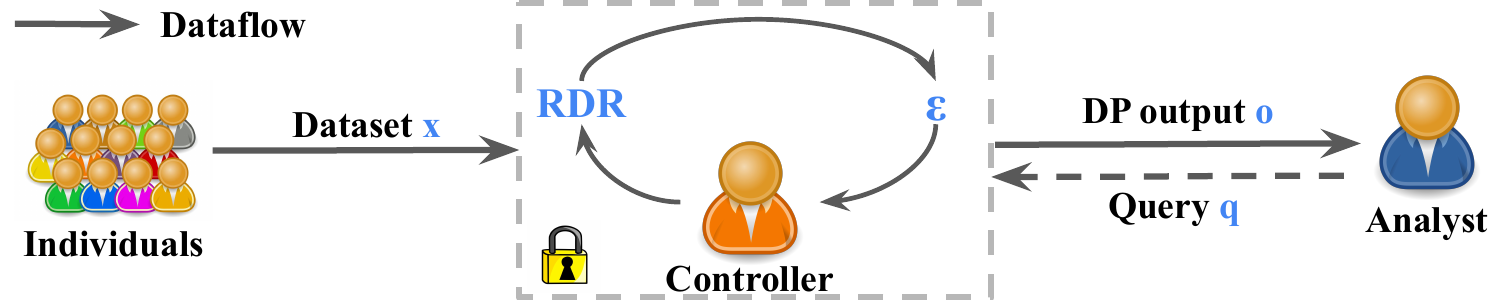}
    \caption{Dataflow of \texttt{Find-$\epsilon$-from-RDR} Algorithm}
    \label{fig:dataflow}
\end{figure}

\mypar{Dataflows and leakage} It is critically important to understand what information is released to whom. Figure \ref{fig:dataflow} shows the involved dataflows in a scenario where a controller uses the \texttt{Find-$\epsilon$-from-RDR} algorithm to find $\epsilon$ and release query output to the analyst. Individuals send their data to the controller; this dataflow remains unchanged from the standard central DP deployment shown in Figure~\ref{fig:agents}. The analyst submits a query to the controller. The controller expresses their privacy preference in terms of $\tau_p$ (and not in terms of $\epsilon$) and runs the algorithm to release an output produced by a DP mechanism (Laplace Mechanism or Gaussian Mechanism in our work) with an $\epsilon$ selected by the algorithm. At this point, no information has been returned to the analyst yet. The RDRs of within-dataset individuals are \emph{never} released to the analyst. The DP output is only released with the controller's approval. The chosen $\epsilon$ is also \emph{not} released since $\epsilon$ is chosen based on data-dependent RDRs, and releasing it directly jeopardizes the DP guarantee. In Section~\ref{sec:algo_ext}, we explain a modification of the algorithm that permits finding and releasing $\epsilon$ while satisfying DP.

\myparnoperiod{Claim:} \textbf{The \texttt{Find-$\epsilon$-from-RDR} algorithm satisfies $(\epsilon, \delta)$-DP}. The algorithm only releases the output $o$ computed using an $(\epsilon, \delta)$-DP mechanism ($\delta = 0$ when $\mathcal{M}$ is Laplace Mechanism), where $\epsilon$ is found by the algorithm that satisfies the controller's privacy preference. If no suitable $\epsilon$ is found, the algorithm does not release anything; in other words, $\epsilon = 0$. Therefore, the algorithm satisfies $(\epsilon, \delta)$-DP.

\mypar{Algorithm complexity and optimizations} The algorithm has a worst-case time complexity of $O(n \lvert \mathcal{E} \rvert)$, where $n$ is the number of individual records in the dataset. Computing the RDRs requires computing the per-instance sensitivity first. When the number of individual records is high, this becomes a performance bottleneck: using memoization saves computation if the number of unique projected records is smaller than $n$. In addition, the computation of each individual's per-instance sensitivity can be parallelized since they are independent of one another, \ie the problem is embarrassingly parallel. Thus, we implement parallelism and compute the per-instance sensitivity for a chunk of records in each process. We show in Section \ref{sec:scalability} that the algorithm is both efficient and scalable.



%% file: sections/section5.tex
\section{Releasing Epsilon Privately and Bounding Privacy Loss Across Queries} \label{sec:algo_ext}

In Section~\ref{sec:release_eps}, we propose a modification of the \texttt{Find-$\epsilon$-from-RDR} algorithm that permits finding and releasing $\epsilon$ while preserving DP. In Section~\ref{sec:3.3bound}, we discuss how the new algorithms answer multiple queries without the need to set a total privacy budget, and we provide the end-to-end DP guarantee in Section~\ref{sec:3.4dp_guarantee}.

\subsection{Finding and Releasing $\epsilon$ using SVT} \label{sec:release_eps}

Since \texttt{Find-$\epsilon$-from-RDR} finds $\epsilon$ in a data-dependent manner, releasing the chosen $\epsilon$ directly may reveal additional information about the within-dataset individuals. Therefore, a modification of the algorithm is needed to preserve the DP guarantee if the controller wants to release both $\epsilon$ and DP output $o$ to the analyst.

\mypar{Key Idea} The \texttt{Find-$\epsilon$-from-RDR} algorithm iterates through candidate $\epsilon$ values and selects the first one satisfying the controller's privacy preference, \ie $f(RDR) \ge \tau$. Our key idea is to make this selection process itself differentially private. We leverage the Sparse Vector Technique (SVT)~\cite{dwork2009complexity} to test candidate $\epsilon$ values and release the first value such that $f(RDR)$ \emph{approximately} passes the threshold, thereby ensuring releasing $\epsilon$ preserves DP. 

\begin{algorithm}[h]
\scriptsize
\SetKw{Continue}{continue}
\SetKwInOut{Input}{Input}
\SetKwInOut{Output}{Output}
\Input{Dataset $x$; a stream of SVT queries $\mathcal{Q}_{svt} = q_1, q_2 \cdots$ with global sensitivity at most $\Delta_{svt}$; threshold $\tau$; privacy budget $\epsilon_{svt}'$ and $\epsilon_{svt}''$} 
\Output{A stream of answers $a_1, a_2 \cdots$ where each $a_i \in \{\top, \bot\}$}
\BlankLine

$\rho = Lap\left(\frac{\Delta_{svt}}{\epsilon_{svt}'}\right)$ \label{algo:svt:line1}

\For{$q_i \in \mathcal{Q}_{svt}$}{ 
    
    \If{$q_i(x) + Lap\left(\frac{2\Delta_{svt}}{\epsilon_{svt}''}\right) \geq \tau + \rho$}{  \label{algo:svt:line3}       
        Output $a_i = \top$ \\
        \Return
    } \Else {
        Output $a_i = \bot$
    }
}
\caption{SVT algorithm proposed by~\citeauthor{lyu2016understanding}~\cite{lyu2016understanding}}
\label{alg:svt}
\end{algorithm}

\mypar{The SVT algorithm} Algorithm~\ref{alg:svt} shows the standard SVT algorithm~\footnote{The algorithm from ~\cite{lyu2016understanding} has an additional parameter $c$ (the maximum number of queries the algorithm can answer $\top$). We set $c = 1$, since we only need one positive answer for our modified algorithm.} proposed by ~\citeauthor{lyu2016understanding}~\cite{lyu2016understanding}, which we will incorporate into our algorithm. SVT takes a dataset, a stream of queries (each with a global sensitivity of at most $\Delta_{svt}$), threshold $\tau$, and privacy parameter $\epsilon_{svt}$ that is divided into $\epsilon_{svt}'$ and $\epsilon_{svt}''$ (\ie $\epsilon_{svt} = \epsilon_{svt}' + \epsilon_{svt}''$), then releases a stream of answers indicating whether each query output is above or below the threshold. The algorithm works by first initializing a Laplacian noise $\rho$ scaled to $\Delta_{svt} / \epsilon_{svt}'$ (line~\ref{algo:svt:line1}). For each query, it generates another term of Laplacian noise scaled to $2\Delta_{svt} / \epsilon_{svt}''$ and adds the noise to the query output. Then the algorithm compares the noisy query output with the noisy threshold (line~\ref{algo:svt:line3}); if it is above or equal to the threshold, the algorithm outputs $\top$ indicating the query output \emph{approximately} passes the threshold, otherwise, it outputs $\bot$. 
Overall, the algorithm satisfies $\epsilon_{svt}$-DP. 

\mypar{A Low-Sensitivity SVT Query} Intuitively, the controller's preference function becomes the SVT query. While the \texttt{Find-$\epsilon$-from-RDR} algorithm offers the controller the flexibility to define any preference functions, many functions have high global sensitivity, making them unsuitable for SVT. For example, the function $RDR_{min}/RDR_{max}$ has a sensitivity of 1. This means to apply SVT, the algorithm will add a Laplacian noise scaled to $1 / \epsilon_{svt}'$ to $\tau_p$ and another Laplacian noise scaled to $2 / \epsilon_{svt}''$ to $RDR_{min}/RDR_{max}$. Clearly, when $\epsilon_{svt}$ is small (\eg less than 1), the threshold will be too noisy and make the algorithm useless for finding $\epsilon$. 


To address this caveat, we propose an SVT query based on the \emph{variance of normalized RDRs}. We first normalize each within-dataset individual's RDR, $RDR_i$, by a global, data-independent RDR upper bound, $RDR_{max}^*$, which depends only on the query's global sensitivity, output dimensionality $k$, and the candidate $\epsilon$ value. Based on Lemmas~\ref{lemma:pri_lap} and~\ref{lemma:pri_gau}, $RDR_{max}^* = \Delta_1 + k\Delta_1/\epsilon$ if Laplace Mechanism is used, and $RDR_{max}^* = \sqrt{\Delta_2^2 + k \sigma^2}$ if Gaussian Mechanism is used. Then the normalized RDR, $RDR_i'$, for a candidate $\epsilon$ is $RDR_i'(\epsilon) = RDR_i(\epsilon) / RDR_{max}^*(\epsilon)$.  

\begin{definition}[\textbf{SVT query}] The SVT query for each candidate $\epsilon$ is defined as:
    \[    
    q_{svt}(\epsilon) = Var(\{RDR_i'(\epsilon) | i \in [n]\}), 
    \] where $n$ is the number of within-dataset individuals.
\end{definition}

The SVT query has a global sensitivity $\Delta_{svt} = 1 / n$, since $0 \le RDR_i'(\epsilon) \le 1$. This means the noise added by SVT will be very small for reasonably large datasets.



\begin{algorithm}[h]
\scriptsize
\SetKw{Continue}{continue}
\SetKwInOut{Input}{Input}
\SetKwInOut{Output}{Output}
\Input{DP mechanism $\mathcal{M}$, dataset $x$, query $q$, threshold $\tau_{var}$, set of candidate $\epsilon$ values $\mathcal{E}$, failure parameter $\delta$, SVT budget $\epsilon_{svt}$} 
\Output{Chosen $\epsilon$, DP output $o$}
\BlankLine

$\epsilon_{svt}', \epsilon_{svt}'' \leftarrow \textsc{assign-svt-budget}(\epsilon_{svt})$ \label{algo2:svt_p1_start} \\


$\rho = Lap\left(\frac{\Delta_{svt}}{\epsilon_{svt}'}\right)$ \label{algo2:svt_p1_end} \\

$x' \leftarrow \textsc{project-query-attributes}(x, q)$ \label{algo2:project-attr} \\

$PIS \leftarrow map()$ \label{algo2:create-map} \\ 

\For{$x_i \in x'$}{
    \If{$x_i \notin PIS$}{
        $PIS[x_i] \leftarrow \textsc{compute-per-instance-sensitivity}(x', x_i, q, M)$
    } 
} \label{algo2:compute-pis}

\tcc*[h]{{\scriptsize Controllers may observe the computed per-instance sensitivity of each individual first, and set $\tau_{var}$ based on their privacy preferences}}\\

\For{$\epsilon \in \mathcal{E}$}{ \label{algo2:loop}

    $RDR \leftarrow list()$ \label{algo2:create-list} \\
    \For{$x_i \in x'$}{
        $RDR_i \leftarrow \textsc{compute-rdr}(PIS[x_i], M, q, \epsilon, \delta)$ \\
        $RDR.add(RDR_i)$ \label{algo2:compute-rdr}
    }

   

    \tcc*[h]{{\scriptsize Run SVT to find the first $\epsilon$ that approximately passes threshold}} 

    \If{$-q_{svt}(\epsilon) + Lap\left(\frac{2\Delta_{svt}}{\epsilon_{svt}''}\right) \ge -\tau_{var} + \rho$}{ \label{algo2:check-rdr}
    
        $o \leftarrow \mathcal{M}(x', q, \epsilon, \delta)$ \label{algo2:compute-dp-output} \\
        \Return $\epsilon, o$ \label{algo2:return}
    }
}
\Return nil
\caption{Find-and-release-$\epsilon$-from-RDR}
\label{alg:find_eps_dp}
\end{algorithm}

\mypar{The \texttt{Find-and-release-$\epsilon$-from-RDR} algorithm} Algorithm~\ref{alg:find_eps_dp} integrates this SVT-based query. First, the algorithm allocate the SVT budget $\epsilon_{svt}$ ( line~\ref{algo2:svt_p1_start}); one allocation strategy~\cite{lyu2016understanding} is to have $\epsilon_{svt}' : \epsilon_{svt}'' = 1 : 2^{2/3}$ to achieve better accuracy of SVT. Then it iterates over each candidate $\epsilon$ and computes the corresponding RDRs the same way as in Algorithm~\ref{alg:find_eps} (line~\ref{algo2:project-attr} - ~\ref{algo2:compute-rdr}). For each $\epsilon$, it computes the SVT query and tests if the variance is approximately below the threshold $\tau_{var}$ (line~\ref{algo2:check-rdr}). The first $\epsilon$ that passes this noisy test is selected and released along with the final DP output $o$ (line~\ref{algo2:return}).

The main advantage of using SVT is that it always consumes a fixed budget $\epsilon_{svt}$ \emph{regardless of the number of SVT queries} (\ie the number of candidate $\epsilon$ values). Controllers understand the consequence of running SVT since they know exactly how much noise is added to both the SVT query and the threshold. They can decide whether they are comfortable choosing an $\epsilon$ that does not exactly conform to their privacy preferences; if not, they may use a different $\epsilon_{svt}$ or $\tau_{var}$ and rerun the algorithm. 

\myparnoperiod{Claim:} \textbf{The \texttt{Find-and-release-$\epsilon$-from-RDR} algorithm satisfies $(\epsilon+\epsilon_{svt}, \delta)$-DP.} The algorithm can be divided into two steps. In the first step, it applies SVT, which is $\epsilon_{svt}$-DP, to determine whether each candidate $\epsilon$ satisfies the controller's privacy preferences. In the second step, if a suitable $\epsilon$ is found, it uses an $(\epsilon, \delta)$-DP mechanism to compute output $o$, and releases both $\epsilon$ and $o$. Therefore, by \emph{sequential composition}~\cite{dwork2014algorithmic}, the algorithm satisfies $(\epsilon+\epsilon_{svt}, \delta)$-DP.

\myparnoperiod{Claim:} Assume the SVT budget $\epsilon_{svt}$ budget is split such that $\epsilon_{svt}' = \epsilon_{svt} / 3$ and $\epsilon_{svt}' = 2 \epsilon_{svt} / 3$. \textbf{For any failure probability $\beta \in (0, 1)$, \texttt{Find-and-release-$\epsilon$-from-RDR} algorithm is $(\alpha, \beta)$-accurate, \\where $\alpha = \frac{6}{n\epsilon_{svt}}\ln(\frac{2|\mathcal{E}|}{\beta})$}; the accuracy of the algorithm refers to the correctness of the chosen $\epsilon$. This means that with probability at least $1 - \beta$, if the algorithm returns a value $\epsilon^*$, then the true variance associated with that value satisfies $q_{svt}(\epsilon^*) \le \tau_{var} + \alpha$. And for any candidate $\epsilon' \in \mathcal{E}$ such that $q_{svt}(\epsilon') < \tau_{var} - \alpha$, the algorithm will not halt at a value $\epsilon < \epsilon'$, thus returning a value at least as large as $\epsilon'$(since candidate $\epsilon$ values are tested in descending order). A complete proof is provided in the Appendix.
 
\subsection{Bounding Privacy Loss Across Queries} \label{sec:3.3bound}

The preceding discussion is valid for one query. In practice, analysts often want to run multiple queries. Combining these queries' outputs will reveal new information, eventually breaking privacy as per the fundamental Law of Information Recovery~\cite{dwork2014algorithmic}. 
In today's deployment of DP, controllers need to set a total privacy budget on the dataset \textit{a priori} and stop accepting queries when the composed $\epsilon$ of all answered queries exceeds the budget. The need for setting a total budget poses a critical challenge for controllers:

\mypar{Controllers do not know the privacy implications of total privacy budget} Similar to choosing $\epsilon$ for a single query, when choosing the total privacy budget, controllers face the same challenge of not knowing the actual privacy implications on the within-dataset individuals after releasing multiple outputs. Oftentimes, they resort to selecting the budget arbitrarily~\cite{dwork2019differential}.  


\mypar{The solution} Instead of requiring controllers to choose a fixed total budget, the algorithm keeps track of the already consumed budget on a dataset, \ie the sum of $\epsilon$ or $\epsilon + \epsilon_{svt}$ of all previously answered queries by \texttt{Find-$\epsilon$-from-RDR} and \texttt{Find-and-release-$\epsilon$-\\from-RDR} respectively; we call it $\epsilon_c$. For an incoming query, the algorithm \emph{truncates} the range of candidate $\epsilon$ values to consider only those larger than $\epsilon_c$. 
The algorithm rejects the query if it fails to find a suitable $\epsilon$ larger than $\epsilon_c$, meaning the controller's privacy preferences cannot be met for any candidate $\epsilon$. 

Our solution creates a \textit{dynamic} bound of total privacy loss. As $\epsilon_c$ keeps growing for each answered query, it becomes more difficult to find a suitable $\epsilon$ that is larger than $\epsilon_c$ since larger $\epsilon$ may not produce RDRs that satisfy controllers' privacy preferences. At each point the algorithm fails to find a suitable $\epsilon$ for a query, the total privacy loss is bounded. The bound is refreshed if a suitable $\epsilon$ is found for a different query. A potential concern is whether there is additional privacy leakage using this dynamic bound. In Section \ref{sec:3.4dp_guarantee}, we show the bound can be formalized as a \emph{privacy odometer} in \emph{fully-adaptive composition}~\cite{rogers2016privacy} and there is no leakage. 

\mypar{Considering analysts' preferences over queries} A natural tension exists between analysts and controllers: analysts want more accurate output, whereas controllers want to protect the within-dataset individuals' privacy by releasing noisy output. Because it gets increasingly harder to find a suitable $\epsilon$ after answering multiple queries as $\epsilon_c$ keeps increasing, analysts may need to prioritize queries of which they need more accurate output. One approach to incorporate analysts' preferences when finding $\epsilon$ is to allow them to assign weights (\ie priority scores) to their queries, then controllers can adjust the threshold $\tau$ based on analysts' preferences, \eg they can set a smaller $\tau_p$ or larger $\tau_{var}$ for high-priority queries so the algorithm can find larger $\epsilon$. Nevertheless, solving related problems, such as how to communicate analysts' preferences to controllers, and how to help controllers decide an appropriate $\tau$ that takes both analysts' accuracy preferences and their own privacy preferences into account, are outside the scope of the paper but remain interesting directions to explore. 


\subsection{End-to-End DP Guarantee} \label{sec:3.4dp_guarantee}

\mypar{Implications of choosing $\epsilon$ \emph{adaptively}} If the algorithm has only answered one query, releasing the query output to the analyst satisfies DP and the total privacy loss is bounded by $\epsilon$ in \texttt{Find-$\epsilon$-from-RDR} and $\epsilon + \epsilon_{svt}$ in \texttt{Find-and-release-$\epsilon$-from-RDR}. However, if the algorithm answered multiple queries (for the same dataset), for each query, $\epsilon$ is chosen \emph{adaptively} because it is chosen based on previously answered queries --- a suitable $\epsilon$ needs to be larger than $\epsilon_c$. Then the total privacy loss is not bounded by any fixed constant since the algorithm does not require setting a total privacy budget. This brings the question of whether releasing multiple query outputs computed using adaptively chosen $\epsilon$ causes additional privacy leakage. 

\mypar{\emph{Privacy odometer} in \emph{fully-adaptive composition}} To answer this question, we construct a \emph{privacy odometer}~\cite{rogers2016privacy}, which provides a probabilistic running upper bound on the realized privacy loss after answering each query. The privacy odometer is proposed to provide privacy guarantees in \emph{fully-adaptive composition}~\cite{rogers2016privacy}, in which the privacy parameters are chosen adaptively based on previous queries. In short, \emph{there is no additional leakage to choose privacy parameters adaptively if the privacy bound is computed using sequential composition}. Therefore, the privacy odometer of \texttt{Find-$\epsilon$-from-RDR} under sequential composition~\cite{dwork2014algorithmic} is defined as:

\begin{definition}[\textbf{Privacy Odometer of \texttt{Find-$\epsilon$-from-RDR}}] Given dataset $x$, for a sequence of $m$ queries answered by \texttt{Find-$\epsilon$-from-RDR} using a fixed failure parameter $\delta$, from which the $i$-th query selects $\epsilon^{(i)}$, the algorithm's \emph{privacy odometer} is the function $\texttt{COMP}_{\delta_g}(\cdot)$, where 
\begin{equation*}
  \texttt{COMP}_{\delta_g}(\epsilon^{(1)},\\ \cdots, \epsilon^{(m)}) =
    \begin{cases}
        \sum_{i = 1}^{i = m} \epsilon^{(i)} &  \text{if $\delta_g \geq \sum_{i = 1}^{i = m} \delta$} \\
        \infty & \text{otherwise}
    \end{cases}       
\end{equation*}
\end{definition}

The privacy odometer updates its running upper bound on privacy loss to $\epsilon_c = \sum_{i = 1}^{i = m} \epsilon^{(i)}$ after the algorithm answers all $m$ queries, and releasing the composed output on the same dataset $x$ satisfies $(\epsilon_c, \delta_g)$-DP. The privacy odometer for \texttt{Find-and-release-$\epsilon$-from-\\RDR} can be defined using the same principle:

\begin{definition}[\textbf{Privacy Odometer of \textsf{Find-and-release-$\epsilon$-from-\\RDR}}] Given dataset $x$, for a sequence of $m$ queries answered by \textsf{Find-and-release-$\epsilon$-from-RDR} using a fixed failure parameter $\delta$, from which the $i$-th query uses a fixed SVT budget $\epsilon_{svt}^{(i)}$ and selects $\epsilon^{(i)}$, the algorithm's \emph{privacy odometer} is the function $\texttt{COMP}_{\delta_g}(\cdot)$, where 
\begin{equation*}
  \texttt{COMP}_{\delta_g}(\epsilon^{(1)},\\ \cdots, \epsilon^{(m)}) =
    \begin{cases}
        \sum_{i = 1}^{i = m} \epsilon_{svt}^{(i)} + \epsilon^{(i)} &  \text{if $\delta_g \geq \sum_{i = 1}^{i = m} \delta$} \\
        \infty & \text{otherwise}
    \end{cases}       
\end{equation*}
\end{definition} 

It is also worth noting that under certain conditions~\cite{whitehouse2023fully}, the privacy odometer achieves the same asymptotic bound as advanced composition~\cite{dwork2010boosting}. For simplicity, we define the privacy odometer of our algorithm under sequential composition instead of advanced composition, and we leave further improvements for future work. 


%% file: sections/evaluation.tex
\section{Evaluation}
\label{sec:evaluation}

We answer the following research questions:

\begin{myitemize}
    \item \textbf{RQ1}: Does presenting the within-dataset individuals' RDRs help controllers choose $\epsilon$? 
    
    \item \textbf{RQ2}: Are our proposed algorithms, \texttt{Find-$\epsilon$-from-RDR} and \\ \texttt{Find-and-release-$\epsilon$-from-RDR} efficient and scalable?
    
    \item \textbf{Microbenchmarks}: What is the effect of varying $\tau_p$ in \texttt{Find-$\epsilon$-\\from-RDR} and $\tau_{var}$ in \texttt{Find-and-release-$\epsilon$-from-RDR}?
\end{myitemize}

\subsection{RQ1: Does RDR Help Controllers Choose $\epsilon$?} \label{sec:5.1RQ1}

To study whether presenting the within-dataset individuals' RDRs helps controllers choose $\epsilon$, we conducted an IRB-approved between-subjects user study where the participants were divided into control and treatment groups. To prevent learning effects, only the treatment group sees the within-dataset individuals' RDRs. We chose to conduct the study in the form of an online survey because our tasks do not require in-person guidance, and it is more convenient for participants to complete the study. In addition, it allows us to collect data from a more representative sample. 

\subsubsection{Recruitment and data collection}

We recruited 100 participants using Prolific~\cite{prolific}. To complete the study, each participant must sign a consent form, which describes the purpose and content of the study, compensation, and data confidentiality. We did not collect identifiable information, and the survey responses were anonymized, meaning the participant's IP address, location data, and contact information were not recorded.

\mypar{Prescreening criteria} When recruiting on Prolific, we prescreened participants based on three criteria: i) they are fluent in English, to avoid confusion due to misunderstanding language; ii) they have completed an undergraduate degree or above; iii) they study Computer Science, Computing (IT) or Mathematics; this is to ensure that the participants have a sufficient quantitative background to complete the study.


\mypar{Tutorial and Screening Quiz} Since the main questions in the study ask the participant to choose $\epsilon$, the participant needs to have basic knowledge about DP and understand the implications of the $\epsilon$ they chose. To ensure each participant knows enough about DP to complete the study, we asked each participant to read a short tutorial about DP at the beginning. This tutorial explains the intuition of DP and the effect of the parameter $\epsilon$ at a high level. Mainly, it explains that DP is generally achieved by adding random noise to the query output, and larger $\epsilon$ means less noise will be added and vice versa. After reading the tutorial, the participant needs to complete a screening quiz that tests their knowledge about the tutorial they just read. Only participants who pass the quiz proceed to conduct the study. Of the 100 participants we recruited, 56 passed the quiz and their responses were used for the study analysis. This quiz ensures that participants have understood the principles of DP, thus strengthening the validity of the user study.

\subsubsection{Study procedure} \label{sec:5.1.2study_proc}



First, the participants were randomly divided into control or treatment groups; 29 participants were in the control group, and 27 were in the treatment group. 
The survey presents a sample from the UCI Adult dataset~\cite{uci-adult} and explains a DP mechanism is used to answer queries on the dataset by adding noise controlled by the parameter $\epsilon$. Then, it presents two queries in random order:
\begin{myitemize}
    \item The number of people with income greater than 50K who are less than 40 years old.
    \item The number of people who have a Master's degree or above and have income $\le$ 50k, grouped by each race.
\end{myitemize}
For each query, the survey asks the participant to choose $\epsilon$ from a slider, with values from 0.5 to 5.0 in an interval of 0.5. Since we want all participants to have a clear task goal, the survey includes a description of the task: choose $\epsilon$ to protect the within-dataset individuals' privacy while yielding as accurate answers as possible.

For the control group, the participant was presented with a table showing the range of possible outputs for each candidate $\epsilon$ (and the raw output when no DP is applied) to indicate the resulting accuracy of each $\epsilon$. The table includes information that any controller can access in a DP deployment today when choosing $\epsilon$. For the treatment group, the participant was presented with the same table, along with the additional information: a plot of the range of within-dataset individuals' RDRs for each candidate $\epsilon$, \ie a box drawn from the minimum to the maximum RDR. An explanation of the RDR was provided to the participants. We also conducted another study in which we recruited a different treatment group and showed the participants RDR histograms instead of range plots. The additional information of RDR is the only difference between the control and treatment groups to ensure we isolate the intervention. Participants in the treatment group were also asked a follow-up question about whether they used the within-dataset individuals' RDRs when choosing $\epsilon$; we included it to ensure that if we observe an effect, we can attribute the effect to the use of RDRs.

\mypar{Remarks} An alternative way to design the intervention for the treatment group is to present participants with the raw RDRs for each candidate $\epsilon$ (along with the per-instance sensitivity of within-dataset individuals), then ask participants to explicitly define a preference function in terms of the RDRs and a threshold to choose $\epsilon$, \ie by running the \texttt{Find-$\epsilon$-from-RDR} algorithm. However, doing so would require recruiting participants who have real-world experiences as data controllers managing private datasets about individuals to devise a privacy preference, who have sufficient quantitative background to examine the per-instance sensitivity and RDR in raw numbers and translate their preference into a preference function, and who have more than a surface-level understanding of DP. In contrast, the participants we recruited come from various backgrounds, and we cannot make those assumptions about them. Instead, our study evaluates whether the visual presentation of RDRs, through range plots or histograms, enables participants to form \emph{implicit} preferences about the risk distribution of the within-dataset individuals and choose $\epsilon$ based on their preferences.    

\subsubsection{Results}

\begin{figure}[htbp]
    \centering
    \includegraphics[width=\linewidth]{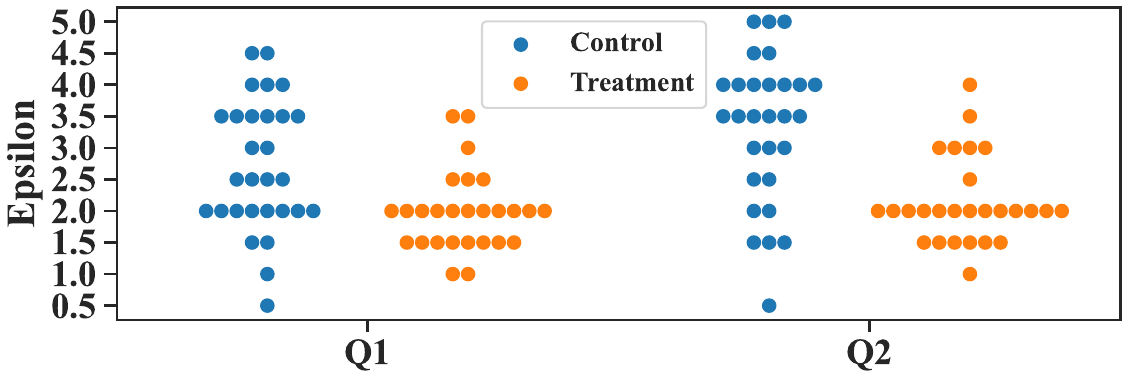}
    \caption{$\epsilon$ chosen by each participant}
    \label{fig:user_study}
\end{figure}


If the intervention (showing the range plots of within-dataset individuals' RDRs) is effective, we expect this would manifest as more consistent answers among participants, since all participants were given the same training and the same task. Therefore, we compare the distribution of $\epsilon$ chosen by participants in the control and treatment groups. Figure \ref{fig:user_study} shows a swarm plot where each point represents the $\epsilon$ chosen by each participant. For both queries, 19 out of 27 participants in the treatment group chose either 2 or 1.5, showing a more consistent answer than the control group. This is because the range of the within-dataset individuals' RDRs when $\epsilon$ is 2 or 1.5 is considerably smaller than that of larger $\epsilon$, and participants know that smaller $\epsilon$ has lower accuracy. This shows that most participants could connect the within-dataset individuals' RDRs and their task goal, and they were able to use the RDRs to choose an $\epsilon$ that satisfies the goal. On the contrary, the $\epsilon$ chosen by participants in the control group are more spread out and have greater variance, suggesting participants shared no general consensus as to what is a good $\epsilon$ to choose. In addition, we ran Mann–Whitney test ~\cite{mcknight2010mann} on the $\epsilon$ chosen from the control and treatment group, with the alternative hypothesis that the two samples come from different distributions. For both queries, we obtained $p < 0.005$. Thus we reject the null hypothesis and conclude that showing the within-dataset individuals' RDRs helped participants choose $\epsilon$.


Lastly, we asked participants in the treatment group if they used the within-dataset individuals RDRs to complete the task: all 27 participants responded Yes, confirming they were indeed ``treated''.


\begin{figure}[h]
    \centering
    \includegraphics[width=\linewidth]{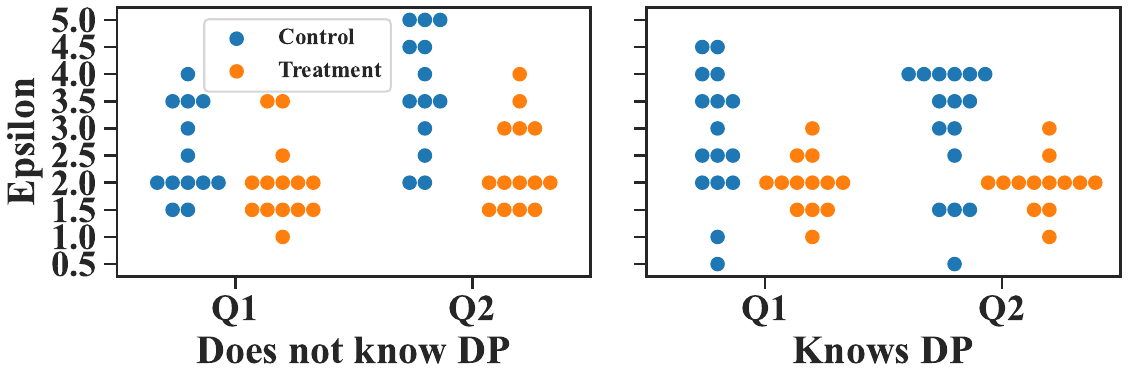}
    \caption{$\epsilon$ chosen by participants who do not know DP (left) and those who know DP (right)}
    \label{fig:user_study_stratify}
\end{figure}


\myparnoperiod{Does knowing DP matter?} Another interesting question is whether the participant's prior familiarity with DP affects how they choose $\epsilon$. At the start of the survey before showing the tutorial, we asked each participant whether they know about DP. 27 participants (out of a total of 56) answered they do not know anything about DP and 29 participants answered they know about DP. Figure \ref{fig:user_study_stratify} shows the $\epsilon$ chosen by participants based on their familiarity of DP. We find similar patterns. Participants in the treatment group tend to choose either 2 or 1.5, regardless of their familiarity with DP, whereas participants in the control group chose $\epsilon$ with greater variance. This suggests participants' prior knowledge of DP does not transfer to making better choices of $\epsilon$; on the other hand, showing the within-dataset individuals RDRs helped participants make more consistent choices whether or not they knew about DP previously.

\myparnoperiod{Range plot vs histogram?} To analyze the effect of different RDR visualizations, we recruited another group of participants - 20 of them passed the quiz, and we asked them to complete the same survey questions as in the previous treatment group, except they were presented with histograms of RDRs instead of range plots.

\begin{figure}[h]
    \centering
    \includegraphics[width=\linewidth]{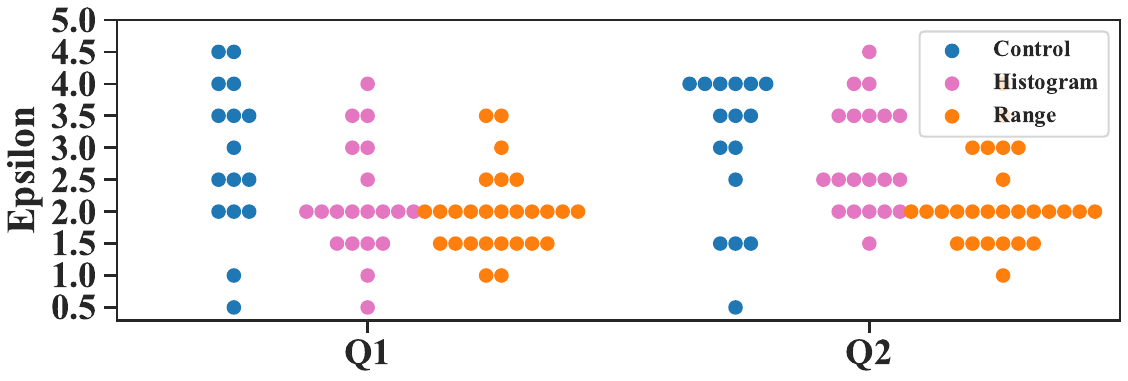}
    \caption{$\epsilon$ chosen by participants who were presented with histograms or range plots, or in the control group}
    \label{fig:user_study_hist_range}
\end{figure}

Figure \ref{fig:user_study_hist_range} shows $\epsilon$ chosen by participants who were presented with histograms or range plots; we include the $\epsilon$ chosen by participants in the previous control group as reference. Participants who saw histograms did not perform better than participants who saw range plots. Especially for Q2, eight participants chose larger $\epsilon$ (3.5, 4, and 4.5). Since the responses are anonymous, we cannot directly ask participants about their decision-making after the survey ends. We deduce one possible explanation might be that for the large $\epsilon$, the majority of RDRs are already small and similar, and only a small portion of RDRs have large values. The participants might think that as long as the privacy of the majority of people is being protected, the task goal is achieved; hence, they chose larger $\epsilon$, so the query outputs are more accurate. However, if the participants only saw the range of RDR, they would be more inclined to choose smaller $\epsilon$ with a smaller range of RDR. Another possible explanation could be that some participants had difficulties interpreting histograms, whereas the range plots are generally much easier to understand. 

The discrepancy of results by tweaking how we present the RDRs to the participants demonstrates that the way we convey information affects their implicit privacy preferences on the within-dataset individuals, which in term affects their decision-making for choosing $\epsilon$. Exploring other tools for presenting and helping data controllers interpret RDR remains an interesting problem.

\subsubsection{Threats to validity} \textbf{External validity:} We recruited participants from diverse backgrounds worldwide via Prolific, with few prescreen criteria so that the participants could complete the study without difficulty. \textbf{Internal validity:} To ensure participants have the minimum knowledge of DP needed to complete the study, each participant must read a tutorial and pass a quiz. Participants can only take the quiz once to minimize the chance of them abusing it if they have multiple chances.
To avoid learning effects, we used a between-subject study to isolate the condition we intended to test. We also randomized the two questions for choosing $\epsilon$.

\subsection{RQ2: Are Both Algorithms Practical?} \label{sec:scalability}

Both \texttt{Find-$\epsilon$-from-RDR} and \texttt{Find-and-release-$\epsilon$-from-RDR} need to be efficient so that it is practical for controllers to use them to find $\epsilon$ for various queries. Thus, we evaluate both algorithms' runtime for different queries and dataset sizes. 

\begin{figure}[h]
    \centering
    \includegraphics[width=\linewidth]{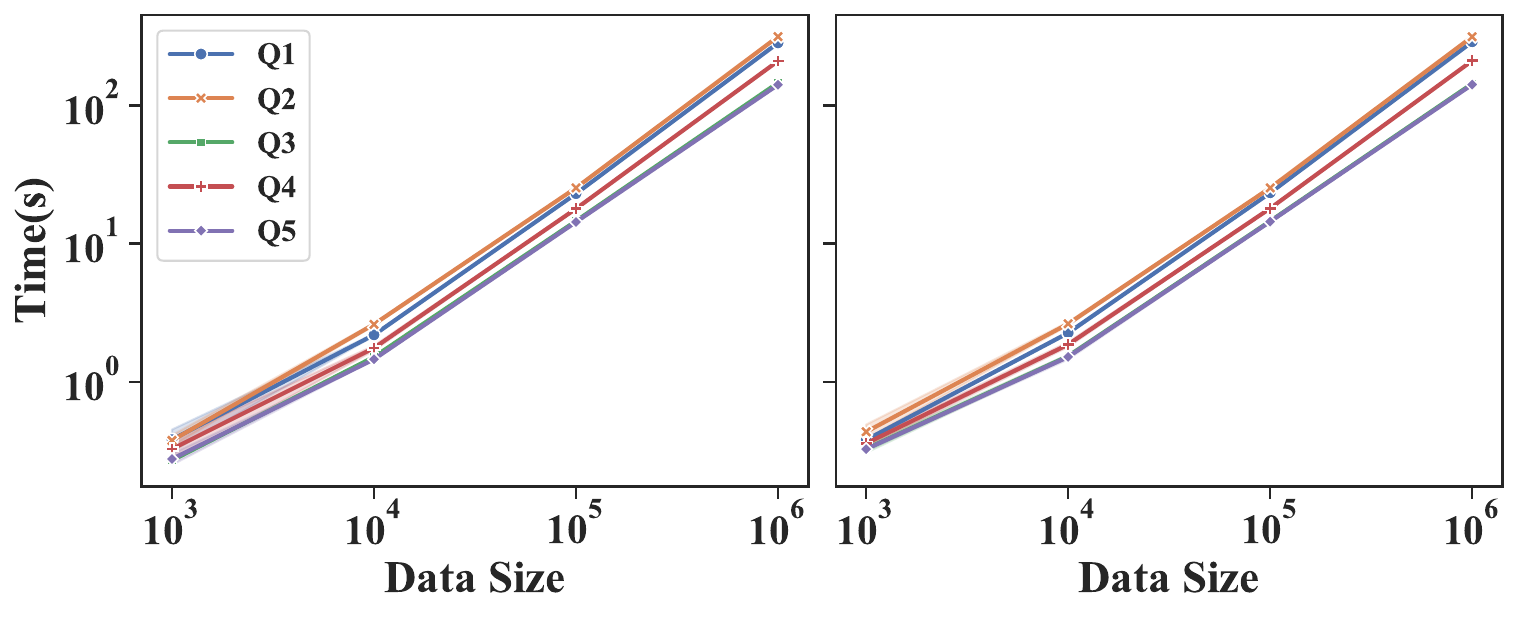}
    \caption{Average runtime (in seconds) of \texttt{Find-$\epsilon$-from-RDR} (left) and \texttt{Find-and-release-$\epsilon$-from-RDR} (right) over 10 runs}
    \label{fig:scalability}
\end{figure}

\begin{figure*}[ht]
    \centering
    \includegraphics[width=\linewidth]{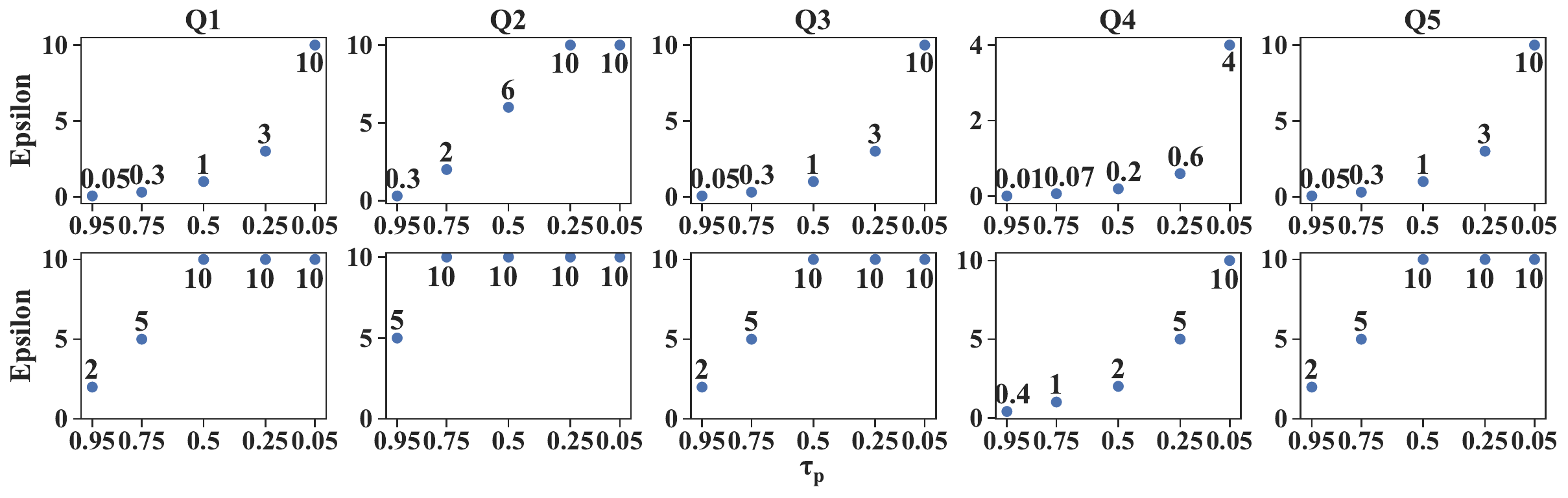}
    \caption{$\epsilon$ found by \texttt{Find-$\epsilon$-from-RDR} under Laplace Mechanism (top) and Gaussian Mechanism (bottom)}
    \label{fig:vary_percentage}
\end{figure*}

\begin{figure*}[ht]
    \centering
    \includegraphics[width=\linewidth]{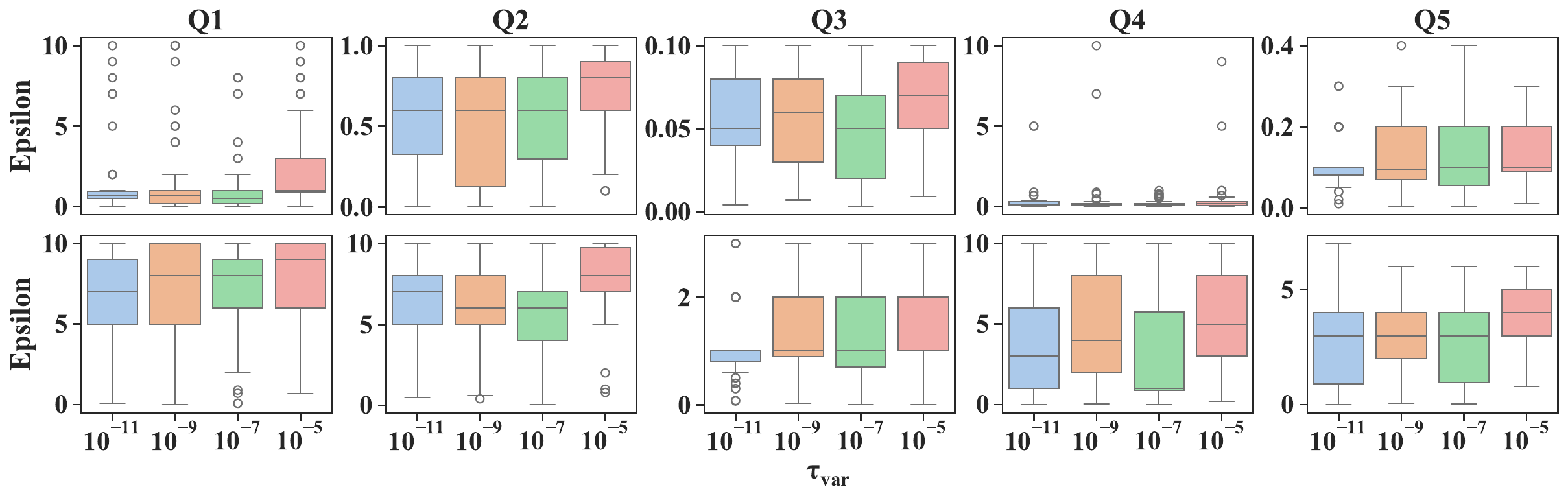}
    \caption{$\epsilon$ found by \texttt{Find-and-release-$\epsilon$-from-RDR} under Laplace Mechanism (top) and Gaussian Mechanism (bottom)}
    \label{fig:vary_svt_threshold}
\end{figure*}

\mypar{Datasets} \label{sec:5.4.1dataset}
We controlled the experiment by generating datasets with 1000, 10000, 100000, and 1 million records from the Adult dataset~\cite{uci-adult}. We sampled records for datasets of sizes (number of records) 1000 and 10000. For datasets of size 100000 and 1 million, we replicated and added random noise to the replicated records to ensure that they were not the same copy of each other. We included all 15 attributes in the dataset. 

\mypar{Queries} \label{sec:5.4.2queries}
We ran both algorithms for the five SQL queries below. We chose a diverse set of queries with attributes of different domain sizes and types. The queries are run independently.
               
\begin{myitemize}
\scriptsize
    \item Q1:\texttt{ SELECT COUNT(*) FROM adult WHERE income=='>50K' AND education\_num==13 AND age==25}
    \item Q2:\texttt{ SELECT marital\_status, COUNT(*) FROM adult WHERE race=='Asian-Pac-Islander' AND 30<=age<=40 GROUP BY marital\_status}
    \item Q3:\texttt{ SELECT COUNT(*) FROM adult WHERE native\_country!='United-States' AND sex =='Female'}
    \item Q4:\texttt{ SELECT AVG(hours\_per\_week) FROM adult WHERE workclass in ('Federal-gov', 'Local-gov', 'State-gov')}
    \item Q5:\texttt{ SELECT SUM(capital\_gain) FROM adult}
\end{myitemize}

\mypar{Parameters} \label{sec:5.4.3params}
We initialize candidate $\epsilon$ values in fixed intervals; specifically, we test $\epsilon$ in $[10, \cdots, 1, 0.9, \cdots, 0.1, 0.09, \cdots, 0.01, 0.099,\\\cdots, 0.001]$ (37 values in total). We use $RDR_{min}/RDR_{max}$ as the preference function and set $\tau_p = 0.95$ in \texttt{Find-$\epsilon$-from-RDR}, and we set $\epsilon_{svt} = 1, \tau_{var} = 10^{-5}$ in \texttt{Find-and-release-$\epsilon$-from-RDR}. We used Laplace Mechanism as the underlying DP mechanism $\mathcal{M}$. We used 8 parallel processes to compute the per-instance sensitivity of each within-dataset individual.

\mypar{Configurations} We ran the experiments on a Chameleon Cloud~\cite{chameleon} instance (Intel Xeon Gold 6242 CPU) with 32 cores and 188GB memory. We used Ubuntu 24.04 and Python 3.12 for all experiments.

\mypar{Results} Figure \ref{fig:scalability} shows two line plots where the x-axis is the data size and the y-axis is the runtime in seconds; both axes are in log scale. The left plot corresponds to the run time of \texttt{Find-$\epsilon$-from-RDR}, and the right plot corresponds to the run time of \texttt{Find-and-release\\-$\epsilon$-from-RDR}. Each line in the plot represents a query, with the error band showing the standard deviation. As shown in both plots, the runtime of both algorithms increases linearly with the data size, and the difference between the two algorithms is negligible. All queries took less than a few minutes to finish; the longest runtime was 5 minutes by Q2 with 1 million records. The runtime is dominated by the time to compute the per-instance sensitivity of each within-dataset individual since this involves first computing the query output on each $x_{-i}$ (the dataset without $i$'s record).

\subsection{Microbenchmarks} \label{sec:microbenchmarks}

\subsubsection{Effect of $\tau_p$ in \texttt{Find-$\epsilon$-from-RDR}} \label{sec:5.3.2}
Given the preference function $RDR_{min} / RDR_{max}$, we test the effect of varying threshold $\tau_p$ and report the $\epsilon$ found. 

\mypar{Setup} We used the original Adult dataset with 48842 rows, and the same five queries and candidate $\epsilon$ values in Section~\ref{sec:scalability}. We run the algorithm by varying $\tau_p = 0.95, 0.75, 0.5, 0.25, 0.05$ and the DP mechanism $\mathcal{M}$ as Laplace Mechanism or Gaussian Mechanism.

\mypar{Results} Figure~\ref{fig:vary_percentage} shows the $\epsilon$ found by \texttt{Find-$\epsilon$-from-RDR} under Laplace Mechanism and Gaussian Mechanism respectively. As shown in both figures, smaller $\tau_p$ leads to larger $\epsilon$, as expected. Using Gaussian Mechanism as the underlying mechanism to compute RDR caused the algorithm to find larger $\epsilon$ compared to using Laplace Mechanism, because with the same $\epsilon$, Gaussian Mechanism is more likely to produce noisier output than the Laplace Mechanism and RDRs that are closer.
Overall, the results empirically demonstrate that the algorithm is useful for finding a suitable $\epsilon$ consistent with the controller's privacy preferences.



\subsubsection{Effect of $\tau_{var}$ in \texttt{Find-and-release-$\epsilon$-from-RDR}} \label{sec:algo2_mircrobenchmark}

Due to the randomness of SVT, we evaluate the effect of varying $\tau_{var}$ by running the algorithm 50 times and report $\epsilon$ found in each round.

\mypar{Setup} We use the same dataset, queries, and candidate $\epsilon$ values in Section~\ref{sec:5.3.2}. We fix $\epsilon_{svt} = 1$. We run the algorithm by varying $\tau_{var} = 10^{-11}, 10^{-9}, 10^{-7}, 10^{-5}$ and the DP mechanism $\mathcal{M}$ as Laplace Mechanism or Gaussian Mechanism.  

\mypar{Results} Figure~\ref{fig:vary_svt_threshold} shows the boxplots of $\epsilon$ found by \texttt{Find-and-\\release-$\epsilon$-from-RDR} under Laplace Mechanism and Gaussian Mechanism respectively. In both figures, for each query, the ranges of $\epsilon$ found are mostly similar for different $\tau_{var}$ except in some cases when $\tau_{var} = 10^{-5}$ and larger $\epsilon$ are found. In general, controllers should still observe the per-instance sensitivity of within-dataset individuals to choose an appropriate $\tau_{var}$. And similarly to what we observed when running \texttt{Find-$\epsilon$-from-RDR}, the algorithm tends to find larger $\epsilon$ using Gaussian Mechanism compared to using Laplace Mechanism. Because of the randomness introduced by SVT, the spread of $\epsilon$ found by the algorithm can cover the entire range of candidate $\epsilon$ values, thus controllers may want to rerun the algorithm multiple times (or with different $\tau_{var}$ and $\epsilon_{svt}$) to observe the effect of SVT and decide which $\epsilon$ to release.

%% file: sections/related_work.tex
\section{Related Work}
\label{sec:relatedwork}


\mypar{Privacy loss accounting} Various methods have been proposed to account for privacy loss in different granularity. Ex-post per-instance DP~\cite{redberg2021privately} is built upon Ex-post DP~\cite{ligett2017accuracy}, which depends on the realized output of a DP mechanism, and Per-instance DP~\cite{wang2019per}, which depends on a fixed dataset and a specific record. Ex-post per-instance DP loss is output-dependent and cannot be used to compare disclosure risks before the query is run. Per-instance DP loss is output-independent but offers no framework to compare risks and choose $\epsilon$. RDR fills this gap and is designed as a relative indicator of disclosure risk on the within-dataset individuals that controllers use to express privacy preferences and choose $\epsilon$. \citeauthor{feldman2021individual}~\cite{feldman2021individual} proposed a privacy filter under Renyi Differential Privacy by tracking the personalized privacy loss for each individual. 
\citeauthor{zhu2022optimal}~\cite{zhu2022optimal} proposed a unified approach for privacy accounting via a characteristic function. \citeauthor{seeman2023privately}~\cite{seeman2023privately} proposed Per Record Differential Privacy, in which only the policy function on each individual’s privacy loss is published but not the actual privacy loss value, and individuals can compute their own privacy loss using the function. Unlike the aforementioned works where the main goal is to provide tighter privacy accounting than standard DP, \emph{RDR is designed to help controllers understand the privacy implications on the within-dataset individuals when choosing $\epsilon$ in standard DP}.  

\mypar{Choosing privacy loss budget} 
One line of work~\cite{murtagh2018usable, nanayakkara2022visualizing, hay2016exploring, thaker2020overlook, john2021decision, nanayakkara2023chances} implements interfaces to visualize the effect of different variables (\eg utility metrics) in relation to $\epsilon$. 
\citeauthor{john2021decision}~\cite{john2021decision} defines and visualizes the data sharing risk. 
Both metrics quantify potential data leakage, but neither metric applies to each within-dataset individual, unlike our RDR.
\citeauthor{ligett2017accuracy}~\cite{ligett2017accuracy} empirically finds the smallest $\epsilon$ consistent with an accuracy requirement in the private empirical risk minimization setting. \citeauthor{ge2019apex}~\cite{ge2019apex} translates a user-specified accuracy bound to a suitable DP mechanism, along with an $\epsilon$ that incurs the least privacy loss while satisfying the accuracy bound. \citeauthor{hsu2014differential}~\cite{hsu2014differential} proposes an economic model for choosing $\epsilon$ by balancing the interests of analysts and prospective participants who will participate if the benefit outweighs the risk. However, this model assumes participants can quantitatively express the "cost" of whether to participate in the study, which is impractical for non-experts.
\citeauthor{kohli2018epsilon}~\cite{kohli2018epsilon} envisions a voting system where individuals submit their preferences over $\epsilon$.
\citeauthor{cummings2021need}~\cite{cummings2021need} and \citeauthor{xiong2020towards}~\cite{xiong2020towards} investigate users' understanding of DP guarantees through user studies.
To the best of our knowledge, \emph{no known approach helps controllers understand the effect of $\epsilon$ on each within-dataset individual's privacy and uses the information to design an algorithm that finds $\epsilon$ automatically, as we do.}

\mypar{Customizing Privacy} Variants of DP have been proposed that allow individuals to specify their own privacy preferences. \citeauthor{jorgensen2015conservative}~\cite{jorgensen2015conservative} proposes Personalized Differential Privacy, in which each within-dataset individual specifies their own privacy preference value instead of using a global privacy loss budget. \citeauthor{he2014blowfish}~\cite{he2014blowfish} proposes Blowfish Privacy, in which data publishers specify their privacy policies in addition to specifying $\epsilon$. Both works extend the standard differential privacy framework and require individuals to define their privacy preferences. Instead, \emph{RDR is only used by data controllers, who are responsible for protecting individuals' data and they can now express their privacy preferences in terms of RDRs.}.


%% file: sections/conclusion.tex
\section{Conclusions} 
\label{sec:conclusions}

In this paper, we showed that by presenting additional information (the RDR) on the relative disclosure risk of the within-dataset individuals when applying DP, controllers gain more understanding of the effect of choosing $\epsilon$. We leverage the RDR to design an algorithm, \texttt{Find-$\epsilon$-from-RDR}, which finds $\epsilon$ based on controllers' privacy preferences in terms of the RDRs of the within-dataset individuals. In addition, we propose an alternative algorithm, called \texttt{Find-and-release-$\epsilon$-from-RDR}, which finds and releases $\epsilon$ privately. We present our solution to bound total privacy leakage when answering multiple queries without requiring controllers to set a total privacy budget. Overall, our work reduces the burden for controllers to choose $\epsilon$ in practice.


%% file: sections/appendix.tex
\appendix

\section{Appendix A: Derivation of RDR Upper Bounds (Lemmas 3.1 and 3.2)}

\begin{lemma} $RDR_i$ upper bound under Laplace Mechanism is
    \[
      \lVert q(x) - q(x_{-i}) \rVert_1 + k \Delta_1 / \epsilon.
    \]
\end{lemma}

\begin{proof}
    
Let $o = q(x) + Z$, where $Z = (Z_1, \dots, Z_k)$ and $Z_j \sim Lap(b)$ are i.i.d. with $b = \Delta_1 / \epsilon$. Let $C_i = q(x) - q(x_{-i})$. 

By Definition 1 and 5 in the paper, $$RDR_i = \mathbb{E}_{o \sim \mathcal{M}_{Lap}(x,q,\epsilon)}\left[\|o - q(x_{-i})\|_1\right] = \mathbb{E}[\|C_i + Z\|_1]$$. 

By the definition of the $\ell_1$ norm and the linearity of expectation: $\mathbb{E}[\|C_i + Z\|_1] = \mathbb{E}\left[\sum\limits_{j=1}^{k} |(C_i)_j + Z_j|\right] = \sum\limits_{j=1}^{k} \mathbb{E}[|(C_i)_j + Z_j|]$.

By the triangle inequality: $\sum\limits_{j=1}^{k} \mathbb{E}[|(C_i)_j + Z_j|] \leq \sum\limits_{j=1}^{k} \mathbb{E}[|(C_i)_j| + |Z_j|]$.

Since $(C_i)_j$ is a constant for each dimension $j$, this simplifies to: $\sum\limits_{j=1}^{k} \mathbb{E}[|(C_i)_j| + |Z_j|] = \left(\sum\limits_{j=1}^{k}|(C_i)_j| \right) + \left(\sum\limits_{j=1}^{k}\mathbb{E}[|Z_j|]\right) = \|C_i\|_1 + \left(\sum\limits_{j=1}^{k}\mathbb{E}[|Z_j|]\right)$.

Since $Z_j \sim Lap(b)$, $\mathbb{E}[|Z_j|] = b = \Delta_1 / \epsilon$. Since all $Z_j$ are i.i.d., $\sum\limits_{j=1}^{k}\mathbb{E}[|Z_j|] = \sum\limits_{j=1}^{k} k\Delta_1 / \epsilon$.

Combining the terms and substituting back $C_i = q(x) - q(x_{-i})$, we get the upper bound: $RDR_i \leq \lVert q(x) - q(x_{-i}) \rVert_1 + k \Delta_1 / \epsilon$.

This completes the proof.

\end{proof}

\begin{lemma} $RDR_i$ upper bound under Gaussian Mechanism is
    \[
     \sqrt{\lVert q(x) - q(x_{-i}) \rVert_2^2 + k \sigma^2}.
    \]
\end{lemma}

\begin{proof}
    
Let $o = q(x) + Z$, where $Z = (Z_1, \dots, Z_k)$ and $Z_j \sim \mathcal{N}(0, \sigma^2)$ are i.i.d. with $\sigma^2 = \frac{2 \Delta_2^2 \ln(1.25/\delta)}{\epsilon^2}$. Let $C_i = q(x) - q(x_{-i})$. 

By Definition 1 and 5 in the paper, $RDR_i = \mathbb{E}_{o \sim \mathcal{M}_{Gau}(x,q,\epsilon, \delta)}[\|o - q(x_{-i})\|_2] = \mathbb{E}[\|C_i + Z\|_2]$. 

By Jensen's inequality, $(RDR_i)^2 = (\mathbb{E}[\|C_i + Z\|_2])^2 \leq \mathbb{E}[\|C_i + Z\|_2^2]$.

By linearity of expectation: $\mathbb{E}[\|C_i + Z\|_2^2] = \mathbb{E} \left[\sum\limits_{j=1}^{k} ((C_i)_j + Z_j)^2\right] = \sum\limits_{j=1}^{k} \mathbb{E}[((C_i)_j + Z_j)^2]$.

Expanding the squared term: $\sum\limits_{j=1}^{k} \mathbb{E}[(C_i)_j^2 + 2(C_i)_j Z_j + Z_j^2] = \sum\limits_{j=1}^{k} (\mathbb{E}[(C_i)_j^2] + 2(C_i)_j\mathbb{E}[Z_j] + \mathbb{E}[Z_j^2])$.

Since $(C_i)_j$ is a constant for each dimension $j$, $\mathbb{E}[(C_i)_j^2] = (C_i)_j^2$. 

Since each $Z_j \sim \mathcal{N}(0, \sigma^2)$, $\mathbb{E}[Z_j] = 0$ and the variance $Var(Z_j) = \mathbb{E}[Z_j^2] - (\mathbb{E}[Z_j])^2 = \sigma^2$, thus $\mathbb{E}[Z_j^2] = \sigma^2$. 

Substituting these values back, we get: $\mathbb{E}[\|C_i + Z\|_2] = \sum\limits_{j=1}^{k} ((C_i)_j^2 + 2(C_i)_j\cdot0 + \sigma^2) = \sum\limits_{j=1}^{k} \left((C_i)_j^2\right) + \sum\limits_{j=1}^{k} \left(\sigma^2\right) = \|C_i\|_2^2 + k\sigma^2$.

We have thus shown that: $(RDR_i)^2 \leq \|C_i\|_2^2 + k\sigma^2$.

Taking the square root of both sides and substituting back $C_i = q(x) - q(x_{-i})$, we get the upper bound: $\sqrt{\lVert q(x) - q(x_{-i}) \rVert_2^2 + k \sigma^2}$.

This completes the proof.

\end{proof}

\section{Appendix B: Accuracy Analysis of \texttt{Find-and-release-$\epsilon$-from-RDR} algorithm}

\myparnoperiod{Claim:} Assume the SVT budget $\epsilon_{svt}$ budget is split such that $\epsilon_{svt}' = \epsilon_{svt} / 3$ and $\epsilon_{svt}' = 2 \epsilon_{svt} / 3$. For any failure probability $\beta \in (0, 1)$, \texttt{Find-and-release-$\epsilon$-from-RDR} algorithm is $(\alpha, \beta)$-accurate, where $\alpha = \frac{6}{n\epsilon_{svt}}\ln(\frac{2|\mathcal{E}|}{\beta})$; the accuracy of the algorithm refers to the correctness of the chosen $\epsilon$. This means that with probability at least $1 - \beta$, if the algorithm returns a value $\epsilon^*$, then the true variance for that value satisfies $q_{svt}(\epsilon) \le \tau_{var} + \alpha$. And for any candidate $\epsilon' \in \mathcal{E}$ such that $q_{svt}(\epsilon') < \tau_{var} - \alpha$, the algorithm will not halt at a value $\epsilon < \epsilon'$, thus returning a value at least as large as $\epsilon'$(since candidate $\epsilon$ values are tested in descending order). 

\begin{proof}
    The SVT mechanism in \texttt{Find-and-release-$\epsilon$-from-RDR} algorithm checks if $-q_{svt}(\epsilon) + Lap\left(\frac{2\Delta_{svt}}{\epsilon_{svt}''}\right) \ge -\tau_{var} + \rho$, where $\rho = Lap\left(\frac{\Delta_{svt}}{\epsilon_{svt}'}\right)$. Let $\nu = Lap\left(\frac{2\Delta_{svt}}{\epsilon_{svt}''}\right)$. Since we assume a budget split of $\epsilon_{svt}' = \epsilon_{svt} / 3$ and $\epsilon_{svt}' = 2 \epsilon_{svt} / 3$, we have $\nu \sim Lap(3\Delta_{svt}/\epsilon_{svt})$ and $\rho \sim Lap(3\Delta_{svt}/\epsilon_{svt})$; let $b = 3\Delta_{svt}/\epsilon_{svt}$. 
    
    Let the total noise on the comparison be $Z = \nu - \rho$. An error occurs if the noise flips the outcome of the comparison, which happens if $|Z| > |q_{svt}(\epsilon) - \tau_{var}|$. We bound the probability that $|Z| > \alpha$ for any of the $\mathcal{E}$ queries. 

    Using a standard tail bound for the difference of two i.i.d. Laplace variables, $\nu, \rho$: $\Pr[|Z| > \alpha] \le 2 \exp (- \frac{\alpha}{2b})$.

    Then we apply a union bound over all $\mathcal{E}$ candidate $\epsilon$ values. The probability that the noise for any query exceeds $\alpha$ is at most: $\Pr[\exists \epsilon \in \mathcal{E}, |Z_\epsilon| > \alpha] \le |\mathcal{E}| \cdot 2 \exp (- \frac{\alpha}{2b})$.

    We set this failure probability to be at most $\beta$: $|\mathcal{E}| \cdot 2 \exp (- \frac{\alpha}{2b}) \le \beta \Rightarrow \alpha \ge 2b \ln (\frac{2|\mathcal{E}|}{\beta})$.

    Substituting $b = 3\Delta_{svt}/\epsilon_{svt}$ and $\Delta_{svt} = 1/n$: $\alpha \ge 2(\frac{3(1/n)}{\epsilon_{svt}})(\ln (\frac{2|\mathcal{E}|}{\beta})) = \frac{6}{n\epsilon_{svt}}\ln(\frac{2|\mathcal{E}|}{\beta})$.

    This establishes the $(\alpha, \beta)$-accuracy guarantee. With probability at least $1 - \beta$, for all $|\mathcal{E}|$ queries, the noisy comparison is within $\alpha$ of the true comparison.
    
\end{proof}